\documentclass[aps,prl,groupedaddress,twocolumn]{revtex/revtex4-2}

\usepackage[dvipsnames]{xcolor}

\usepackage{chemformula}
\usepackage{placeins}
\usepackage{soul}
\usepackage{bm}
\usepackage{amssymb}
\usepackage[pagewise]{lineno}
\usepackage[normalem]{ulem}
\definecolor{tangerine}{rgb}{0.944,0.522,0}
\definecolor{verde}{rgb}{0.,0.6,0}
\definecolor{rosso}{rgb}{0.9,0.0,0.2}
\definecolor{magenta}{rgb}{0.9,0.2,0.9}

\newif\ifhighlight

\newcommand{\highlight}{\highlighttrue}
\highlight
\newcommand{\editor}[2]{%
  \expandafter\newcommand\csname #1note\endcsname[1]{%
    \textcolor{#2}{(\textbf{#1note:} \textsc{##1})}}%
  \expandafter\newcommand\csname #1\endcsname[1]{%
    \ifhighlight\textcolor{#2}{##1} \else ##1\fi}%
  \expandafter\newcommand\csname #1cancel\endcsname[1]{%
    \ifhighlight\textcolor{#2}{\sout{##1}}\fi}%
  \expandafter\newcommand\csname #1change\endcsname[2]{%
    \ifhighlight\textcolor{#2}{\sout{##1} ##2}\else ##2\fi}%
  \newenvironment{#1text}{\ifhighlight\color{#2}\fi}{\color{black}}
}

\editor{resub}{blue}
\editor{DT}{verde}

\setchemformula{radical-radius = .3ex}

\DeclareUnicodeCharacter{03B2}{\ensuremath{\beta}}

\begin{document}

\title{Reconstructions and Dynamics of $\beta$-Lithium Thiophosphate Surfaces}

\author{Hanna Türk}
\affiliation{Laboratory of Computational Science and Modeling, Institut des Mat\'eriaux, \'Ecole Polytechnique F\'ed\'erale de Lausanne, 1015 Lausanne, Switzerland}
\author{Davide Tisi}
\affiliation{Laboratory of Computational Science and Modeling, Institut des Mat\'eriaux, \'Ecole Polytechnique F\'ed\'erale de Lausanne, 1015 Lausanne, Switzerland}
\author{Michele Ceriotti}
\affiliation{Laboratory of Computational Science and Modeling, Institut des Mat\'eriaux, \'Ecole Polytechnique F\'ed\'erale de Lausanne, 1015 Lausanne, Switzerland}
\date{\today}%

\begin{abstract}

Lithium thiophosphate (LPS) is a promising solid electrolyte for next-generation lithium-ion batteries due to its superior energy storage, high ionic conductivity, and low-flammability components. Despite its potential, the high reactivity of LPS with common contaminants such as atmospheric water, preparation solvents, and electrode materials poses significant challenges for commercialization. The lack of understanding regarding the structure, morphology, and chemical behavior of LPS's surface slows down the search for solutions to these issues. 
Here, we utilize a machine learning interatomic potential to achieve a fundamental, atomistic understanding of the mechanical and chemical properties of the β-\ch{Li3PS4} surfaces. Employing molecular dynamics simulations, we identify relevant surface complexions formed by surface reconstructions, determine their surface energies and compute the Wulff shape of $\beta$-LPS. The most stable complexions exhibit properties distinctly different from the bulk, including amorphization, increased density, decreased conductivity and large deformation of the structure building blocks.
We demonstrate that these surfaces are not static, but undergo significant dynamical activity which is clearly identified by an analysis featuring a time-averaged structural descriptor.  
Finally, we examine the changes of the electronic structure induced by the surface complexions, which provides us with details on changes in surface reactivity and active sites, underlining the importance to investigate surface complexions under realistic conditions.

\end{abstract}
\maketitle

\vspace{0.2cm}

%

%
%


%
%
%
%

\clearpage
\section{Introduction}

Solid-state batteries (SSBs) are emerging as a promising advancement in sustainable energy storage, offering significant improvements in safety and performance compared to conventional lithium-ion batteries (LIBs), while still being suitable for the same applications \cite{batesAreSolidstateBatteries2022, liAdvanceReviewSolidstate2021, Stegmaier2022}. Unlike conventional LIBs, which use flammable liquid electrolytes,
SSBs utilize solid-state electrolytes (SSEs), thus minimizing safety hazards by eliminating the possibility of leakage of dangerous substances. 
Potential performance enhancements include increased temperature ranges \cite{kimSolidStateLiMetal2021, batesAreSolidstateBatteries2022}, 
 faster charging without
electrolyte polarization \cite{janekChallengesSpeedingSolidstate2023} due to the higher conductivity of lithium ions in many SSE, as well as possibly higher
energy and power densities by potential usage of Li metal anodes and bipolar stacking \cite{jungSolidStateLithiumBatteries2019}.

Among many available SSEs, lithium thiophosphate (LPS) is notable for its high ionic conductivity \cite{liuAnomalousHighIonic2013a, katoLithiumionconductiveSulfidePolymer2021}, mechanical and electrochemical stability \cite{dewaldExperimentalAssessmentPractical2019, katoHighpowerAllsolidstateBatteries2016},  long life-cycle \cite{manthiramLithiumBatteryChemistries2017} and resource-efficient composition of the earth-abundant elements sulfur and
phosphorous enabling cheap, sustainable applications at large scales \cite{staackeTacklingStructuralComplexity2022a}. 

Despite the many advantages of LPS as SE, 
LIBs containing it are still under development, due to the high reactivity of LPS towards many electrode materials as well as water. This
necessitates handling the material in inert atmospheres with water-free solvents, which significantly increases production costs and hinders commercialization \cite{huoInterfaceDesignEnabling2023a, huangChallengesSolutionsSolidState2024}.

Currently, passivation strategies, such as material doping with e.g. halides \cite{morinoDegradationArgyroditetypeSulfide2023, poudelTransformingLi32024, nikodimosEffectSelectedDopants2022} or softer bases according to the HSAB (hard and soft acids and bases) concept\cite{pearsonHardSoftAcids1963, sahuAirstableHighconductionSolid2014, luAirStabilitySolidState2022, nikodimosMoistureRobustnessLi6PS5Cl2024}, as well as solvent inclusion \cite{luTuningMoistureStability2023a} or
coatings with Li\textsuperscript{+}-conducting materials such as \ch{Al2O3} or polymeric binders \cite{leeSelectionBinderSolvent2017, riphausSlurryBasedProcessingSolid2018, leeThiolEneClick2019, liSulfidebasedCompositeSolid2024, braksInterfacialStabilizationPrelithiated2024}, are being explored.
However, these approaches are also restricted by the
material's high reactivity \cite{huoInterfaceDesignEnabling2023a, liangChallengesInterfaceEngineering2022,liSulfidebasedCompositeSolid2024}.
The lack of understanding of LPS's surface structure, morphology, and chemical behavior slows down the development of this technology, emphasizing the need for fundamental insights into surface reaction mechanisms and interfacial properties. 
Obtaining atomistic insights on LPS is rather challenging, given that LPS is a complex material with multiple possible compositions and crystal structures. We here concentrate on \ch{Li3PS4}, which has 3 crystal polymorphs ($\alpha$ - space group \textit{Cmcm}, $\beta$ - \textit{Pnma}, and $\gamma$ - \textit{Pmn2\textsubscript{1}} \cite{mercierStructureT6trathiophosphateLithium1982, kuduStructuralDetailsLi3PS42022,hommaCrystalStructurePhase2011}) as well as amorphous glassy structures. Sufficiently high ionic conductivity for battery applications has been observed for the $\alpha$, $\beta$, and amorphous structures. Among these, the $\beta$ phase has received a lot of attention by researchers since it was successfully stabilized at room temperature by Liu et al. in 2013 \cite{liuAnomalousHighIonic2013a}. 

In recent years computational studies have addressed the bulk and surface \cite{lepleyStructuresLiMobilities2013, kimStructuralElectronicDescriptors2020, weiReconstructionElectronicProperties2022, maranaComputationalCharacterizationBetaLi3PS42022, gigliMechanismChargeTransport2024, tisiThermalConductivityLi2024, staackeTacklingStructuralComplexity2022a} properties of LPS. The majority of these studies have been limited by the use of empirical potentials \cite{ariga2022new,forrester2022} and \textit{ab-initio} molecular dynamics (AIMD) \cite{de2018analysis}, based on density functional theory (DFT) with generalized gradient approximation (GGA) \cite{perdew1996generalized,perdew1996generalizedHole, senAtomiclevelInsightsHighly2024}. 
The former can provide useful mechanistic insights, but have not yet been able to correctly predict the activation energies of the conductive phases \cite{forrester2022} and are not suitable for a detailed study of surface properties. Quantum mechanical approaches, on the other hand, are more accurate and can give insightful evidence of relevant effects \cite{zhang_targeting_2020, smith_low-temperature_2020,lepleyStructuresLiMobilities2013}, but their high computational cost hinders their applicability to the large systems required to study surface effects, e.g. surface reconstruction and dynamics under realistic conditions. 
To address this cost/accuracy tradeoff one can use machine learning to construct interatomic potentials, which, once properly trained over accurate quantum mechanical data, have the accuracy of DFT at a cost only marginally higher than classical force fields \cite{Deringer2021,WANG2018178,zeng2023deepmdkit,Bartok2010, Behler2007, Smith2017, schutt2022schnetpack, Rupp2012,Butler2018,fourGenBeheler,Unke2019, Batzner_NatCommun_2022_v13_p2453,PhysRevB.104.104309}. 
Such machine learning interatomic potentials (MLIPs) provided already great insights into the properties of bulk LPS \cite{staackeTacklingStructuralComplexity2022a,gigliMechanismChargeTransport2024, tisiThermalConductivityLi2024, nikodimosEffectSelectedDopants2022}. Furthermore, their reduced computational demand enables simulations of larger systems beyond the practical limits of DFT, being able to now adequately investigate the structure of grain boundaries \cite{harperTrackingLiAtoms2025} and helped to uncover the nature of relevant surface complexions of other materials
\cite{dillonComplexionNewConcept2007,Timmermann2020}. 

In the following, we first elucidate the surface morphology of $\beta$-LPS (\ch{Li3PS4}) using a MLIP-driven MD to achieve thorough energetic relaxation. We then characterize these surfaces and find properties that are distinctly different from those of the bulk. Molecular dynamics (MDs) simulations of a reconstructed (100) surface provide insights on the surface stability and dynamics under realistic conditions.
Finally, we take a look at the electronic structure of the low energy complexions, which provides information on the reactivity of the surface. 

\begin{figure}
  \centering
    \includegraphics[width=\linewidth]{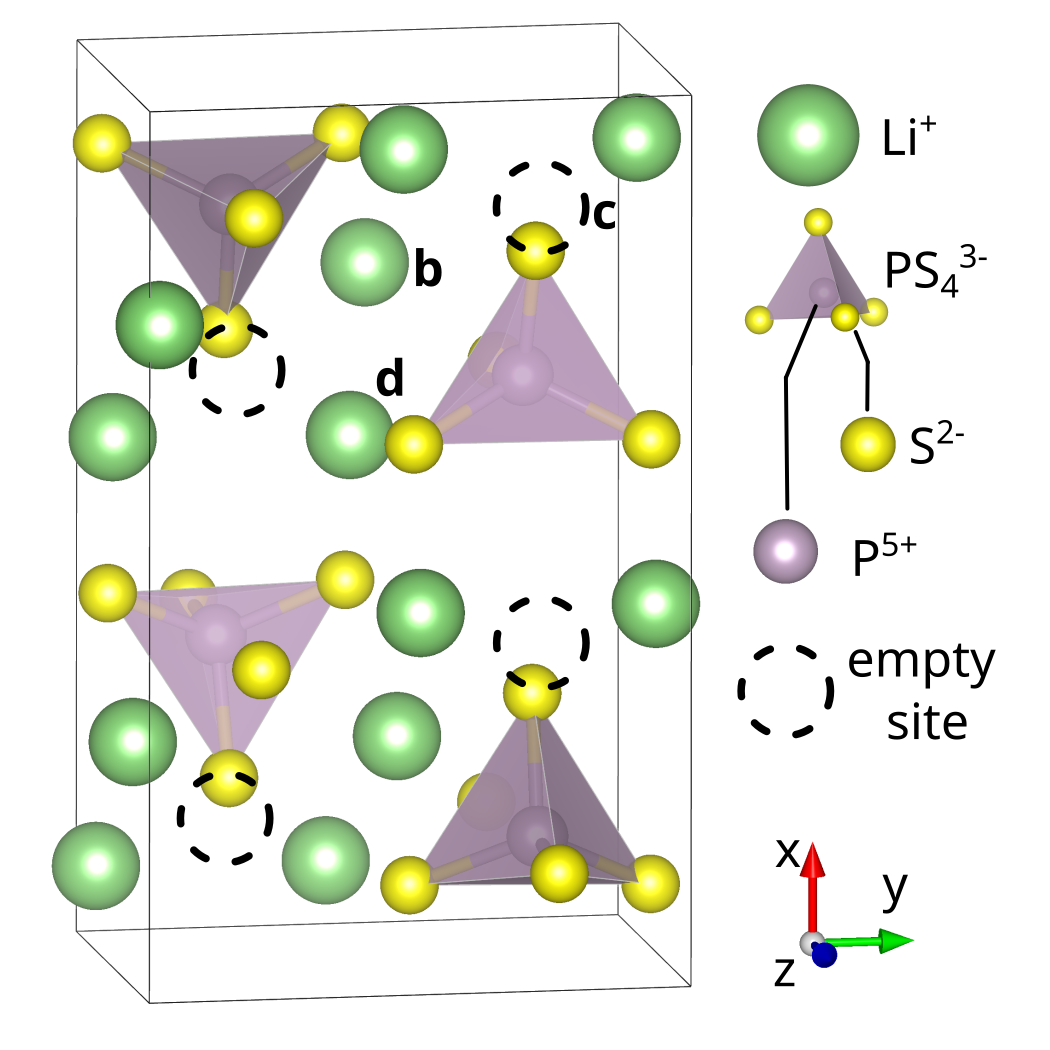}
  \caption{DFT relaxed unit cell of $\beta$-LPS with fully occupied d and b Wyckhoff sites and empty c sites (dashed black circles). d, b, and c Wyckhoff sites according to \cite{lepleyStructuresLiMobilities2013} are labelled.}
  \label{fig:betalps}
\end{figure}

\section{Results and Discussion}

\subsection{Surface Reconstructions and Surface Energies}

$\beta$-LPS has $Pnma$ crystal structure, in which the lithium ions can occupy three different possible lattice sites. These sites are named d, b, and c Wyckoff sites \cite{lepleyStructuresLiMobilities2013}. Their position in a DFT relaxed unit cell of $\beta$-LPS is shown in Figure~\ref{fig:betalps}.
The lattice parameters of the DFT calculation as well as of the fitted MLIP agree reasonably with experimental values (compare Table~\ref{SItab:lattice}, Supporting Information \cite{SubMat}).

\begin{figure*}
  \centering
    \includegraphics[width=1\linewidth]{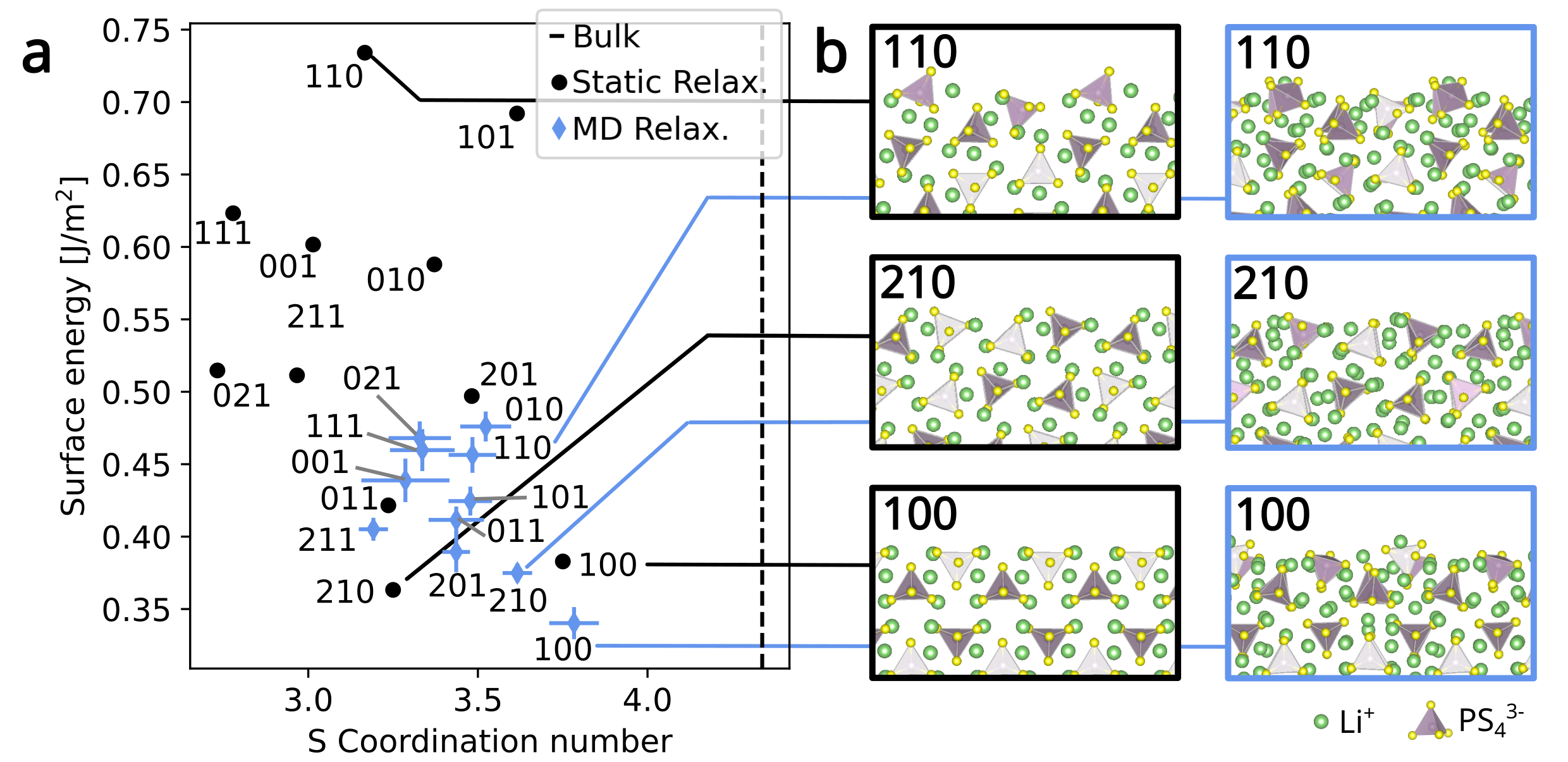}
  \caption{(a) Surface free energy of selected, low Miller-index surfaces %
  of a statical relaxation with the MLIP and the relaxation after a short MD simulation plotted against the sulfur coordination number. For the structures obtained from MD simulations, the energy is averaged over 10 randomly selected and relaxed snapshots along the (minimum 220~$ps$ long) MD and the error bars are their standard deviation. (b) Structures of selected LPS surfaces in yz-plane after statical relaxation (left) and after an MD followed by a relaxation (right). The surface Miller indices are given in the top left of each panel.}
  \label{fig:surface_stability}
\end{figure*}

We use this relaxed unit cell to cut several $\beta$-LPS surfaces with low Miller indices. We then perform molecular dynamics (MD) simulations to determine the surface stabilities under realistic conditions (300~K, ambient pressure). During the simulation, the surfaces undergo reconstructions, resulting in surface structures possessing a lower potential energy. This is illustrated in Figure~\ref{fig:surface_stability}a, where we compare the different LPS surface energies $\gamma$ (see Section~\ref{SIsec:functions}, Supporting Information \cite{SubMat}, for definition) after a static relaxation (black circles), as often performed when surfaces are investigated, with the surfaces we obtained after the MD simulations (blue diamonds). 
Considering the statically relaxed structures, one can see that the surfaces vary greatly in their stability. They follow roughly the correlation of surface energy to the coordination number of the sulfur atoms, which has been described by Kim et al. \cite{kimStructuralElectronicDescriptors2020} (See section~\ref{SIsec:functions}, Supporting Information \cite{SubMat}, for the definition of surface energy and sulfur coordination number).
In comparison, all surfaces that were obtained from an MD simulation and then relaxed have a lower surface energy compared to the statically relaxed structures (except for the (210), which has the same surface energy). 
MD relaxation also changes the order of the most stable surfaces, with the (100) surface being a bit more stable than the (210) surface after reconstruction.
Other surfaces, such as the (110), show a drastic increase in stability.
In general, it can be observed that the reconstructed surfaces are closer to each other in surface energies, clustering at values between 0.35-0.48~$\frac{J}{m^2}$ (see Table~\ref{SItab:surfe}, Supporting Information \cite{SubMat}, for values).
This clustering in energy is accompanied by having also a more similar sulfur ion coordination number, which increases for all surfaces. Similarly, the lithium-ion coordination is also increasing (compare Figure~\ref{fig:SILicoord}, Supporting Information \cite{SubMat}). For many catalytic and electrochemical systems the coordination of surface ions has been correlated to their reactivity, with higher coordinated atoms being less reactive \cite{falicovCorrelationCatalyticActivity1985,coperetSurfaceOrganometallicCoordination2016}. Thus, this provides indications that the reconstructed, relaxed surfaces might be less reactive than the statically relaxed one. However, note that we are currently only considering relaxed structures, and this might change when dynamics are involved (as we will observe later). 

An explanation for the drastic change in surface energies during the MD simulation can be found when observing the structure of the surfaces. 
During the MD simulation, the lithium ions are given the chance to hop between sites, and thus can redistribute to the most energetically favorable configuration.
Note, that this is not only happening at the surface, but also in the bulk. 
At elevated temperatures (637~K), it was experimentally found that the d sites in $\beta$-LPS are fully occupied, whereas the b (70\%) and c (30\%) sites are partially occupied \cite{hommaCrystalStructurePhase2011}. We also find in our simulations that all start from a purely d and b site occupation of lithium ions (as shown in Figure~\ref{fig:betalps}) after a short simulation time the c sites are also partly occupied. 
Visualizing the surface, however, the lithium ions become more randomly distributed, and many of them cannot be clearly assigned to Wyckoff sites anymore. %
Such drastic redistribution that requires not only an escape of the local energy well but also the breaking of the symmetry of the structures cannot be found by static relaxation but is readily obtained with finite-temperature MD equilibration.

For the \ch{PS4}\textsuperscript{3-} tetrahedra, rotations are possible during MD, which cannot only lead to lower surface energies, but also to changes in the surface morphology.
When comparing to the sulfur coordination number, it also becomes obvious that the sulfur ions become more buried in the surface during the reconstruction, as the coordination number increases for the surfaces. 
These changes can clearly be observed in Figure~\ref{fig:surface_stability}b, which compares surface cuts after a static relaxation (left) with the corresponding surface complexions after an MD simulations followed by relaxations (right).
When comparing the structure before and after the MD simulation, one clearly observes a strong restructuring of the surfaces, and a loss of the crystal order of the underlying bulk. As previously discussed, this is especially pronounced for the \ch{Li+} ions, which are now occupying non-clearly defined positions in the crystallographic lattice. Note that the depiction of the reconstructed surfaces are only one possible realization of the structure at this specific Miller index, and dynamic movements of lithium ions as well as (as we will see later) tetrahedra rotations drive a dynamic evolution of the surface structure over time. %

At this point it is noteworthy that the surfaces of LPS are not uniquely defined by their Miller indices, meaning that adding or omitting some \ch{PS4}$^{3-}$ or lithium ions of a specific surface does not alter the Miller index. Thus, for one Miller index, multiple possible surface terminations exist which differ in e.g. lithium or sulfur richness \cite{maranaStabilityFormationLi32023}. However, as we observed in this study, the naturally occurring surface reconstructions are governed by ion diffusion and redistribution. Hence, different surfaces with the same Miller index are energetically very similar after an MD simulation. As a result, the surfaces have comparable properties almost regardless of their initial surface termination (compare section~Figure~\ref{fig:SI111}, Supporting Information \cite{SubMat}). 

\begin{figure}
  \centering
    \includegraphics[width=\linewidth]{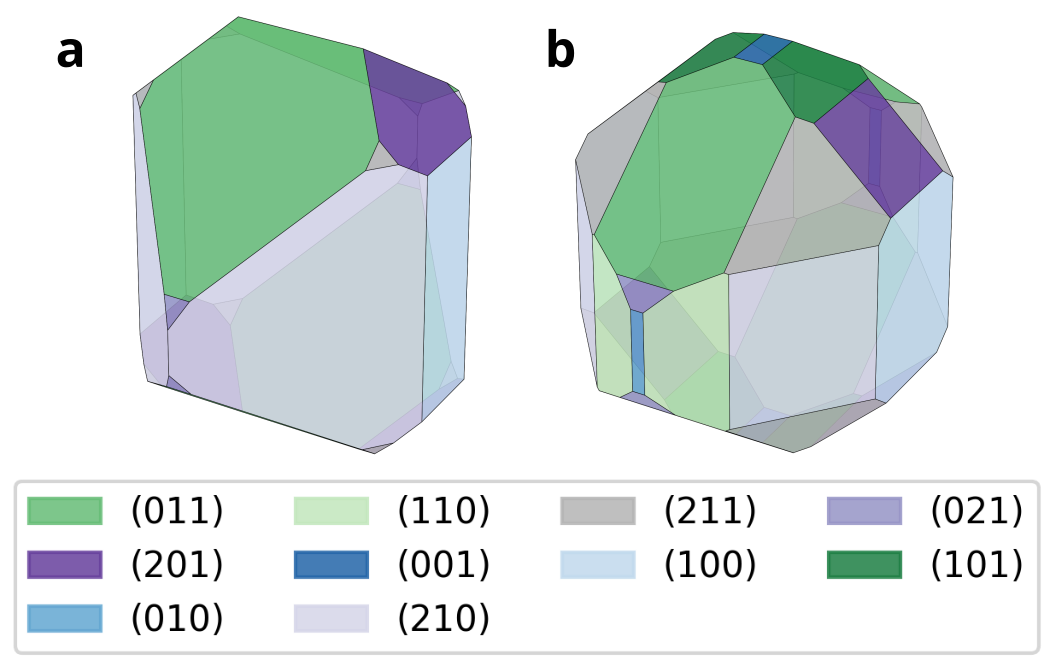}
  \caption{Wulff constructions of LPS after (a) statical relaxation and (b) after an MD that allowed for surface reconstructions.}
  \label{fig:wulff}
\end{figure}

\subsubsection{Wulff Construction}
We computed the Wulff shapes for both the statically relaxed and reconstructed surfaces. Figure~\ref{fig:wulff} presents the resulting Wulff constructions (see Figure~\ref{SIfig:wulff} for comparison with previously reported Wulff shapes). In the case of the statically relaxed structure (Figure~4a), the (210) and (011) surfaces dominate.
In contrast, the Wulff shape of the reconstructed surfaces differs significantly, incorporating a greater variety of surface orientations and appearing more spherical overall. This is expected, as all the surface complexions exhibit comparable surface energies, preventing any single surface from dominating.

\subsection{Structural Characterization of the Surface Complexion}

As we have seen in the previous section, the reconstructed surfaces are energetically more favorable than the statically relaxed structures. 
In this section we analyze in more detail the nature of the surface reconstruction. 

\begin{figure*}
  \centering
    \includegraphics[width=1\linewidth]{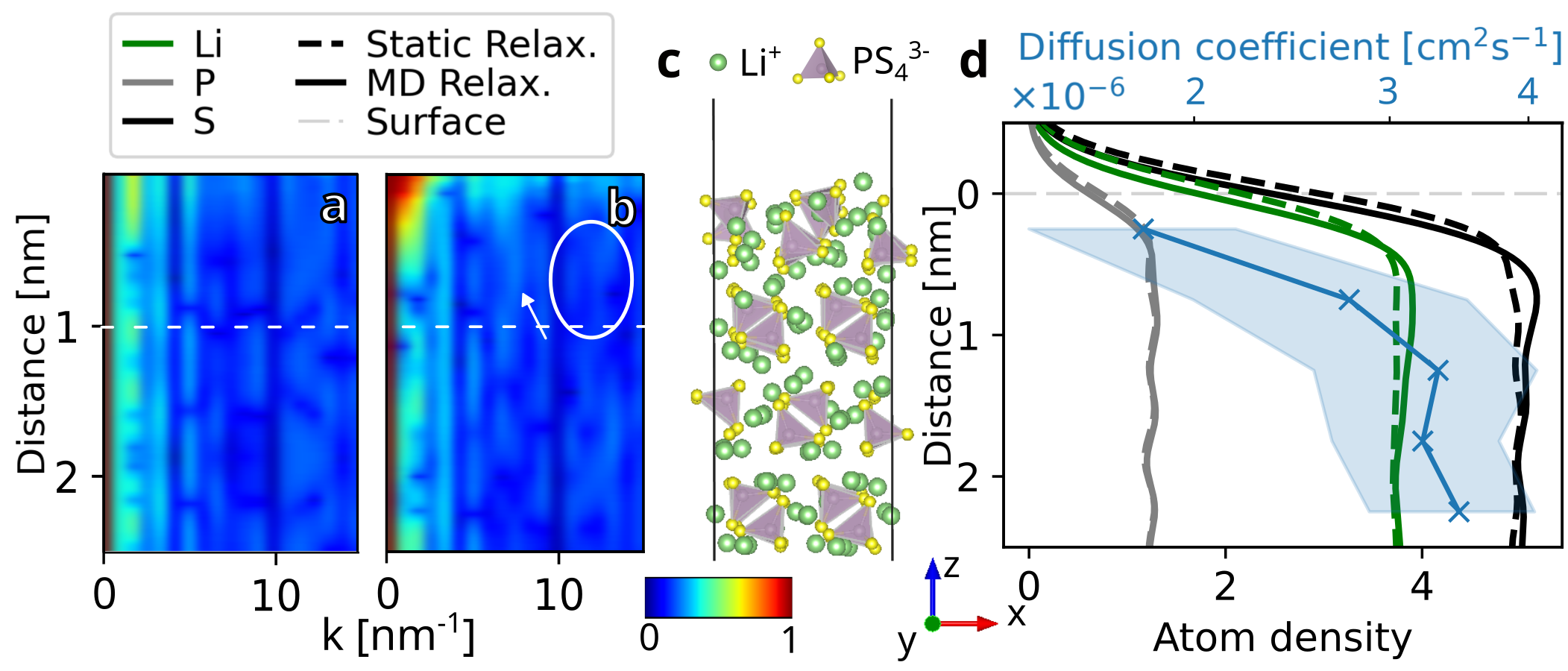}
  \caption{Changed Properties of the (100) surfaces after the formation of the complexion. A partial Fourier Transform of the density of a cell with (100) surface at z=0 of the relaxed cell of a) the static relaxation and b) after an MD and relaxation. The density mimicked by overlaying each atom with a normalized Gaussian, see the methods section for details. Areas remarkably different in the surface area in comparison to (a) are marked with a white arrow or are encircled. c) Image of the surface aligned with the vertical axis in (a), (b) and (d). d) The ionic diffusion coefficient (blue) from MD simulations for the reconstructed surface on a logarithmic scale with the standard deviation of the three MDs per surface displayed as envelope.
  Also, (d) depicts the element resolved densities (green, gray, black) after a MD simulation followed by a relaxation (solid) compared to a static relaxation (dashed).}%
  \label{fig:surface_properties}
\end{figure*}

Figures~\ref{fig:surface_properties}a and b show the 
spatially resolved 1D Fast Fourier Transforms of a (100) surface without (a) and with (b) an MD simulation before the relaxation. In these semi-reciprocal plots, the vertical axis aligns with the (100) surface cell depicted in Figure~\ref{fig:surface_properties}c and the density shown in Figure~\ref{fig:surface_properties}d, which corresponds to the z-direction in the coordinate system and with the Gibbs dividing surface corresponding to the zero of the scale. The horizontal axis shows the Fourier Transformation of the atom density of the cell in panel c at the corresponding distance from the surface.

In Figure~\ref{fig:surface_properties}a, which was computed from the statically relaxed structure, the index reflections do not change significantly towards the surface, indicating that the order of ions is preserved. This, however, changes in panel~b, which shows the partial Fourier transform of the reconstructed (100) surface. 
Although the persistence of low index reflections confirms that long-range order is still present, the short-range order of the crystal changes towards the surface. Firstly, there is a 
shift of the reflection at $k$=6.8~$nm^{-1}$ to 7.6~$nm^{-1}$ (marked with an arrow). This, together with the loss of signals at high $k$ (encircled area), indicates the loss of high index reflections due to the reduction of short range order. 
Hence, the crystal amorphizes towards the surface, with the \ch{PS4}\textsuperscript{3-} tetrahedra remaining in place, though partly rotated. Furthermore, the lithium ions are not at clearly defined lattice sites anymore. This also explains the higher similarity of surface energies for the different reconstructed surfaces, as the partly amorphized surface complexions are able to 
reorient high energy surface features and 
are thus less dependent on the Miller index of the surface and the underlying bulk crystal orientation. 
Based on the distance-resolved profiles in Figure~\ref{fig:surface_properties}b, one can observe that the loss of short-range order associated with the surface reconstruction extends about 1~nm into the surface (indicated by a dashed white line). 
Figure~\ref{fig:surface_properties}d shows that the
atom density of the cell increases towards the surface. The amorphization thus also leads to a partial intrusion into the empty sites in the LPS crystal, leading to a denser packing.
Figure~\ref{fig:surface_properties}d also shows the diffusion coefficient (blue line) of the lithium ions computed in 5~\r{A} thick slices of the $z$-axis of the surface slab. The middle of each slice is used as the $z$-axis value, e.g. the data point at 2.5~\r{A} covers the distance 0-5~\r{A} from the surface in the $z$-direction. The diffusion decreases towards the surface, which is most likely related to the increased density, as the surface collapse during reconstruction intrudes into the empty sites and thus reduces the number of jump destinations for the ions, which diffuse by hopping between nearest-neighbor
vacancy sites in bulk LPS \cite{lepleyStructuresLiMobilities2013}. We found this decrease of ionic conductivity for most investigated surfaces and it seems to be independent of the diffusion direction (see Figures~\ref{SIfig:diff}). Notably, surfaces such as (210), which exhibit lower structural collapse and small energy change after MD relaxation (see Fig.~\ref{fig:surface_properties}), tend to show a smaller reduction in surface diffusion coefficients. For more detail, see \ref{SIfig:diffxyz} in Supporting Information \cite{SubMat}.

The observed surface reconstruction meets perfectly the criteria to classify as a surface complexion \cite{dillonComplexionNewConcept2007,Luo2019}, which feature typically (partial) amorphization, a thickness of around 1~$nm$, and distinctly different properties relative to the bulk.
Complexions have been shown to be the most stable surface termination for other materials \cite{Luo2019,Timmermann2020} as well as some material interfaces \cite{Gotsch2021, turkComplexionsElectrolyteElectrode2021, Tuerk2022, Stegmaier2022, turkBoonBaneLocal2024}.
 Experimentally, highly surface sensitive methods are needed to resolve the surface structure and its properties, because the surface complexion is very thin. 
\subsection{Dynamical Behavior of the Surface Complexion}

\begin{figure*}
  \centering
    \includegraphics[width=1\linewidth]{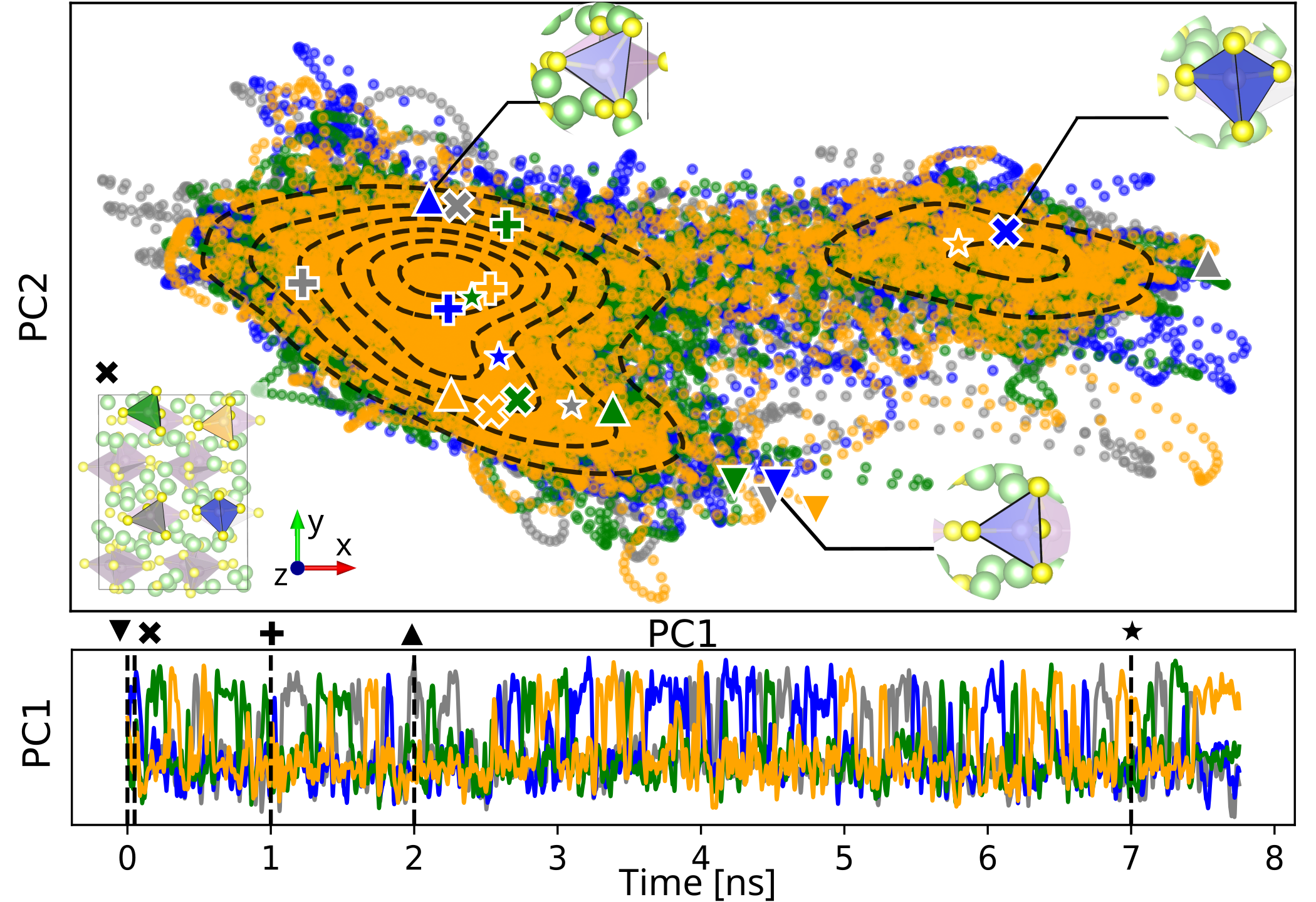}
  \caption{Principal component analysis of the atomic environments of the phosphorous atoms with time averaged SOAP vectors, during one 7.75~$ns$ long MD simulation at 500~$K$ of a (100) surface (top). The trajectories of the different tetrahedra are visualized in gray, blue, green, and orange (compare their respective location in the top view of the surface at the bottom left).  The black dashed lines represent the level curves, showing the position of the two minima. Certain points in time are indicated with special markers ($\blacktriangledown$, $\boldsymbol{\times}$, $\boldsymbol{+}$, $\blacktriangle$, and $\bigstar$) corresponding to the initial surface structure obtained from a DFT relaxation ($\blacktriangledown$), as well as tetrahedra environments at 0.05~$ns$, 1~$ns$, 2~$ns$, and 7~$ns$ of the molecular dynamics simulation cf. the corresponding markers in the bottom panel. The temporal evolution of the PC1 is shown (bottom). } 
  \label{fig:soaps}
\end{figure*}

As we already discussed, room temperature simulations of the complexion show a dynamical evolution of the structure. In order to characterize this further, we analyze the behavior of the phosphorous-centerd atomic environments, using SOAP features to capture the spatial distribution of S, P, and Li neighbors.
In order to de-emphasize short-time local fluctuations, and inspired by approaches that identify long-time behavior of structural descriptors \cite{kelchnerDislocationNucleationDefect1998,stukowskiAutomatedIdentificationIndexing2012,carusoTimeSOAPTracking2023,crippaDetectingDynamicDomains2023,carusoClassificationSpatiotemporalCorrelation2024,donkorLocalStructuresCritical2024, cioniSamplingRealTimeAtomic2024},
we perform a moving average of the descriptors over the trajectory,
\begin{align}
    \bar{\zeta} (A_i, t) = \int ds \zeta (A_i, s) g(s-t) .
\end{align}
The resulting time-averaged descriptor $\bar{\zeta}$ for the atom $i$ in structure $A$ at the time $t$ is thus obtained by integrating the descriptor in a time interval discretized in $s$ and broadened by a Gaussian $g$.

We then perform a principal component analysis (PCA) of the time-averaged descriptors.
The results of this analysis are shown in Figure~\ref{fig:soaps} (top). In the Figure, separate colors are used for each of the four surface tetrahedra, and their individual positions in the simulation cell are displayed with the used color-coding in the left inset of the depicted principal component analysis.

The PCA shows that there are two distinct tetrahedra states in the complexion, with one larger state on the left corresponding to tetrahedra exposing one plane to the surface, and a more localized state on the top right of the PCA that consists of tetrahedra with an exposed edge on the surface. These states are clearly separated by the first principal component (PC1), which correlates with the overall coordination number of the phosphorous within a cutoff of 5.5~\textit{\r{A}} (compare Figure~\ref{fig:SIpc1coord}, Supporting Information \cite{SubMat}). However, the coordination number offers a less clear-cut separation of the two states (compare Figure~\ref{fig:SIpc_coordination}). The correlation of PC1 with the ionic coordination of the phosphorous roots from the redistribution of the surrounding atoms, and especially lithium ions, according to surface tetrahedra alignment, resulting in a higher coordination and slight 'burying' of the edge-exposing tetrahedra. 
This correlation also explains the wider spread of the left cluster in the PCA, which includes surface tetrahedra with its exposed plane completely parallel with the surface (left top of PCA in Figure~\ref{fig:soaps}, compare round inset in the top middle), as well as tetrahedra with exposed planes that are more tilted in respect to the surface (bottom middle area of Figure~\ref{fig:soaps} and the corresponding round inset) and thus more 'buried' into the surface.

The temporal evolution of PC1, as depicted in the bottom of Figure~\ref{fig:soaps}, shows rapid switches of the tetrahedra between the two states with lifetimes in the order of around 100~$ps$. Notably, always just one tetrahedron is switching to this state, showing that there is a coupling between neighboring tetrahedra. 
This correlated flipping aligns very well with the mechanism of transitions between different polymorphs \cite{gigliMechanismChargeTransport2024}. 
Nevertheless, to rule out finite size effects from this observation, we performed similar simulations for larger atomistic cells, which showed still active switching between the surface states (compare Section~\ref{SIsec:dynamics}, Supporting Information \cite{SubMat}).
Note that these simulations were carried out at 500~$K$, however, a conservative estimation of the reaction rate at 300~$K$ shows that this is still a process that takes place on a time scale of less than 200~$ms$
(compare Section~\ref{SIsec:dynamics}, Supporting Information \cite{SubMat}). 
Thus, the surface is highly dynamic at operating conditions and a correct description of its reactivity must include these surface reconstruction phenomena. To obtain an indication of the effect of these dynamic structural changes on the reactivity of the surface, we performed electronic structure calculations.

\subsection{Electronic Structure of the Surfaces}

\begin{figure*}
  \centering
    \includegraphics[width=1\linewidth]{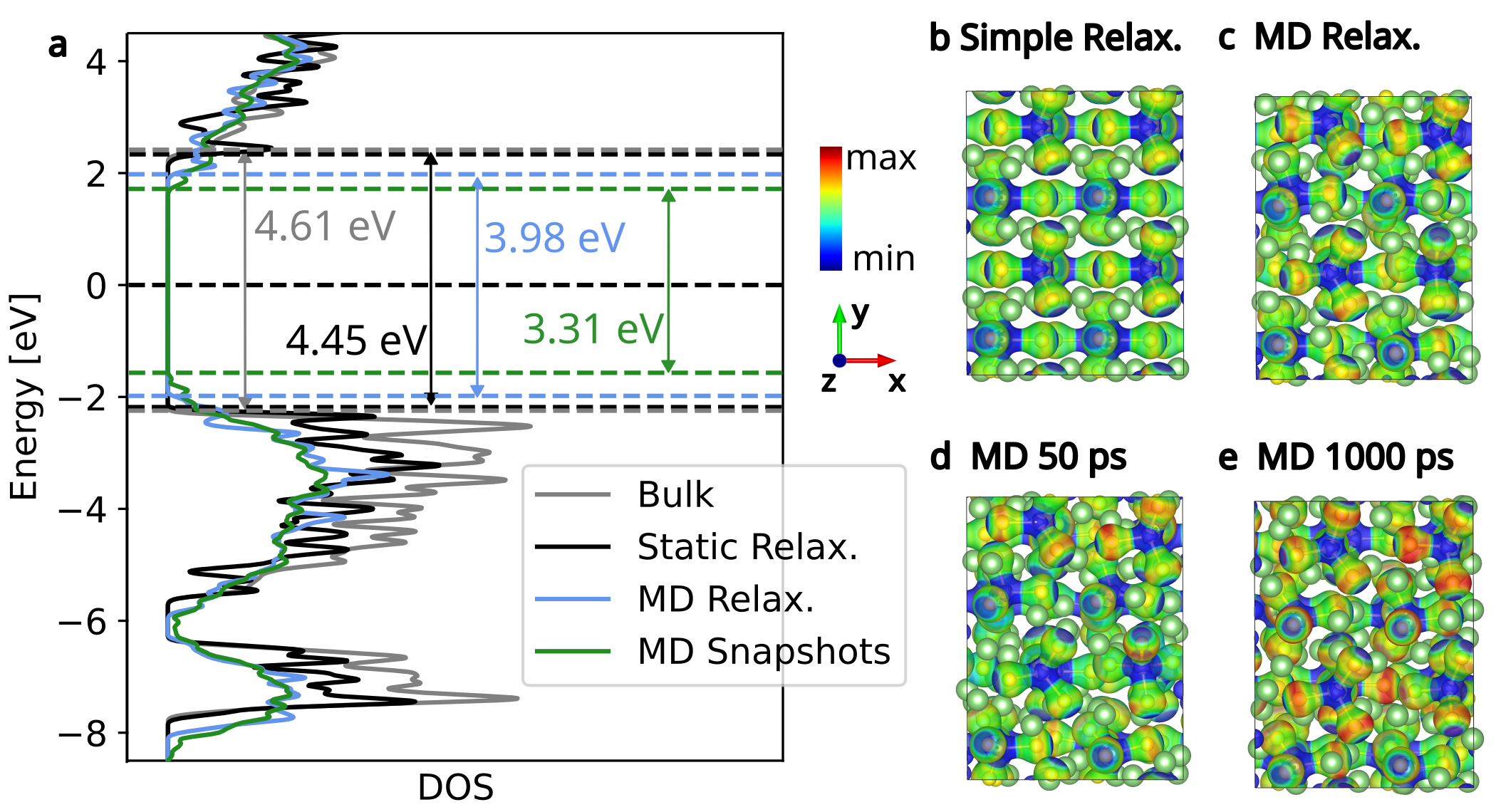}
  \caption{Electronic structure of the (100) surface complexion. a) Density of state of LPS bulk, a statically relaxed structure, an MD relaxed structure with a rotated tetrahedron, and snapshots from an MD simulation. Top view on the electrostatic potential computed with DFT of a (100) surface for (b) the statically relaxed structure, (c) a relaxed MD structure with rotated tetrahedron taken after 50~$ps$, as well as (d) and (e) unrelaxed MD snapshots at 50~$ps$ and 1000~$ps$. The used isosurface level was 0.074~$a_0^{-3}$ and electrostatic potentials were aligned so that the electrostatic potential of the fixed atom layer aligns (compare Figure~\ref{fig:electrostatalign}, Supporting Information \cite{SubMat}).}
  \label{fig:dos}
\end{figure*}

Figure~\ref{fig:dos}a shows the density of states (DOS) of the relaxed bulk structure, the statically relaxed and MD relaxed surfaces, as well as the averaged DOS of four surface structures taken directly from the MD simulation at 500~$K$ without relaxation, which is closer to the realistic operation conditions of LiBs.
The DOS of the bulk structure has the largest band gap ( 4.61~$eV$), which is slightly narrowed to 4.45~$eV$ at the statically relaxed surface and reduced further to 3.98~$eV$ for the MD relaxed structure. For the structures taken directly from the MD simulation without relaxation, the band gap is renormalized to 3.31~$eV$, as a consequence of thermal disorder.

The partial density of states (pDOS) reveals that for all structures the highest occupied states are the sulfur p-states, whereas the lowest unoccupied states consist of sulfur p- and d-states  (compare Figures~\ref{fig:SIpdos_bulk}, Supporting Information \cite{SubMat}).

Figure~\ref{fig:dos}b-e illustrate the electrostatic potential of the (b) statically relaxed, the (c) MD relaxed surface and (d)/(e) at unrelaxed MD snapshots in the same potential range for the top view of a (100) surface (see Figures~\ref{fig:SIsimple}, \ref{fig:SImdrelax}, \ref{fig:SImd500}, and \ref{fig:SImd10000}, Supporting Information \cite{SubMat} for other views). %
The statically relaxed structure has a homogeneous distribution of the electrostatic potential at the surface, in which all \ch{PS4}\textsuperscript{3-} tetrahedra have the same maximum that scarcely diverges from the bulk below. In contrast, the electostatic potential of the MD relaxed structure varies more between the different tetrahedra and individual sulfur atoms.
This effect is even more pronounced for the MD snapshots, where the electrostatic potential shows local fluctuations that also change between the snapshots.
Overall, the structure appears to be very dynamical, with constant alteration of the surface moieties and thus reactive sites. 
The broader range of local potentials of such a dynamic surface will obviously have implications for the study of the surface reactivity. 
However, to quantify the impact of the surface dynamics and the concomitant fluctuations of the electostatic potential on the surface reactivity, a more detailed analysis of relevant molecules and reaction pathways is needed. 

\section{Conclusion}

Leveraging the recent advances in molecular simulation methods, we were able to uncover the surface complexions of $\beta$-LPS under realistic conditions. 
Our investigations show that the surfaces reconstruct 
when a temperature is applied, leading to marked differences in structure and surface energies compared to static geometry relaxation, which retains much of the initialized symmetry.

We were thus able to identify states with a lower surface energy for each investigated Miller index (except (210), which did not show substantial relaxation). These new surfaces are more similar in energy, which influences the Wulff shape to be more round compared to Wulff shapes previously reported for this material, which were based on simple DFT relaxations.
Furthermore, these newly identified surface reconstructions do not only feature different properties compared to the bulk including higher density, lower conductivity, partial amorphization, and narrower band gap, but also structural defects associated with the rotation and reorientation of \ch{PS4} tetrahedra. 

By analyzing MD trajectories in terms of time-averaged SOAP features, we could reveal the presence of clearly distinct states for the tetrahedra, as well as their dynamics. 
The surface complexion is dynamical at room temperature, and constantly moving between different surface states. 
Finally, a further investigation of the surface complexion revealed large fluctuations of the electrostatic potential during dynamics, leading to a broader range of local potentials.

Overall, this study demonstrates the importance of sampling surface reconstructions in realistic thermodynamic conditions, to accurately determine structure, energetics, and electronic properties.
These observations have obvious implications for the study of surface reactivity, as well as cathode-electrolyte interfaces, although an explicit study of surface reactivity requires further extending the chemical space covered by our machine learning potential, as well as access to longer timescales to sample dynamics and capture rare and activated events. Exploring the surface's reactivity with water and searching for molecules that can stabilize LPS by passivating reactive surface sites will be subject to future work. 

\section{Methods}

\subsection{Dataset generation}
Structures were manipulated using the Atomic Simulation Environment (ASE) 3.22 \cite{hjorthlarsenAtomicSimulationEnvironment2017a}. Surfaces are generated with pymatgen 2024.11.13 \cite{tranInhibitionEpidermalGrowth2016}, such that no P-S bonds were broken and two sets of slabs, one with a minimum of 2~$nm$ and one with a 5~$nm$ thickness were obtained. The thinner set was used to generate the training data, the larger one to investigate surface reconstructions with MD (compare below).

To create an initial set of training structures of the 2~$nm$ surfaces, we performed GEN1-xTB \cite{grimmeRobustAccurateTightBinding2017} calculations with DFTB+ \cite{hourahineDFTBSoftwarePackage2020} and i-PI \cite{litmanIPI30Flexible2024}. These were later complemented by MD simulations with an earlier
version of our MLIP model. For all MD simulations, the most diverse were selected with
farthest point sampling from scikit-matter\cite{goscinskiScikitmatterSuiteGeneralisable2023b}, computed with DFT and added to the training set. 

Our generated training set for LPS surfaces for all three crystal polymorphs ($\alpha$, $\beta$, and
 $\gamma$)
and relevant surface adsorbates (water and H$_2$S). To be precise, the dataset
consists of a total of 6244 structures, of these are 2432 bulk, 657 plain surfaces, 2140 possible adsorbates such as water and H$_2$S as well as isolated \ch{PS4} tetrahedra, 851 surfaces with adsorbates and 164
Li$_3$PS\textsubscript{3-x}O\textsubscript{x} with 0.75$\leq$x$\leq$3. 

\subsection{Density Functional Theory Calculations}
DFT calculations were performed using the Vienna Ab initio Simulation Package (VASP 6.4.1) \cite{kresseEfficientIterativeSchemes1996a} with PBE pseudopotentials (semicore p and s states are considered valence states for Li, hard potentials for P and S).
The used functional was PBEsol \cite{perdew1996generalized} for relaxations performed at the DFT level and the creation of the dataset. PBE0~\cite{adamo1999toward} was used for the analysis of the electronic structure. For the surfaces, a dipole correction perpendicular to the surface slab and the Grimme dispersion corrections with zero-damping function were used. The cells that were used to compute the density of states and the electrostatic potential all contain 72 Li, 24 P, and 96 S atoms (compare Figure~\ref{fig:dos}b and c). For the geometry optimization, the lowest 5~\r{A} of the structure were kept fixed to mimic bulk-like behavior. For the DOS of the MD snapshots, we added and normalized the DOS of all four snapshots. Full images of the surfaces can be found in Figure~\ref{fig:SIsimple}, \ref{fig:SImdrelax}, \ref{fig:SImd500}, and \ref{fig:SImd10000}, Supporting Information \cite{SubMat}.
For the calculation of the partial DOS, we projected on spheres centered on the individual atoms with the Shannon ionic radius of the corresponding element \cite{Shannon1976}. For the electrostatic potential, only Hartree and ionic potential were taken into account. The electrostatic potentials were aligned such that the vacuum level of the fixed bulk structure matches (compare Figure~\ref{fig:electrostatalign}, Supporting Information \cite{SubMat}). For the visualization of the electrostatic potential, we used a density isosurface level of 0.074 a$_0^{-3}$ and electrostatic potentials
on the isosurface ranging from $-$8 to $-$3.65~$eV$. The visualization was done with VESTA 3.5.8 \cite{mommaVESTA3Threedimensional2011}. The corresponding pDOS were visualized with sumo \cite{mganoseSumoCommandlineTools2018}.

\subsection{MLIP fitting}
We trained a 
Point Edge Transformer (PET) \cite{pozdnyakovSmoothExactRotational2023} implemented in 
metatrain \cite{METATRAIN}. 
For training, we used a training:validation:test set split of 8:1:1. The resulting errors on the full test set were 3.9~$meV$/atom for the energies and 83.7~$meV$/\textit{\AA}  for the forces. 
When only considering test structures containing Li, P and S (=305 structures), an even lower error of 2.9~$meV$/atom for energies and 54.5~$meV$/\textit{\AA}\textbf{} for forces was observed (compare Figure~\ref{SIfig:rmse}, Supporting Information). 
The transformer was trained with 2 graph neural network layers, 2 transformer layers of size 256 and head size 8, and a cutoff of 5~\textit{\r{A}}.
The initial learning rate was 5$\cdot 10^{-5}$, the batch size 50, and the energy weight for the loss of 0.05 computed for each atom.
We used a ZBL potential for short range interactions.

\subsection{Atom Density, Ionic Conductivity and Fast Fourier Transform}

To mimic the atomic density, the atomic positions were overlayed with Gaussian functions in 3D space. For Figure~\ref{fig:surface_properties}d, the density was summed over the y-direction to give a representitive picture of the density of the full cell in this 2D image.
For the semi-reciprocal plots (Figure~\ref{fig:surface_properties}a and b), the respective 3D atomic density obtained as described above was computed for each z value binned in 0.1~\textit{\AA} bins. Each of these slices in z-direction was further sliced in y-direction and the FFT was computed (for only this one slice in y-direction to achieve higher resolution) to form a spatially resolved array of FFT signals. Atomic structures were visualized using VESTA~3.5.8 \cite{mommaVESTA3Threedimensional2011}.
To assess the ionic conductivity, we performed MD simulations of 500~$ps$ with a time step of 1~$fs$ at NVT (500~$K$). All cells were previously to this equilibrated under NPT conditions at 500~$K$ and 1~$bar$ for 20~$ps$. To estimate the error of the calculations, we performed three independent calculations for each surface.

\subsection{Wulff Constructions}

Wulff constructions were calculated using WulffPack \cite{rahmWulffPackPythonPackage2020}. %

\subsection{\label{sec:methodsmd} Molecular dynamics simulations}
To calculate the surface reconstructions, we performed MD calculations of the 5~$nm$ thick surfaces obtained with pymatgen. Simulations were performed in LAMMPS \cite{LAMMPS2022} with the PET MLIP via the metatensor calculator  \cite{METATRAIN}.
First, we slowly equilibrated the cells for 35~$ps$ at 1.013~$bar$ with an slowly increasing temperature from 0.1-300~$K$ with a timestep of 0.5~$fs$. Then, we ran a NVT ensemble calculation at 300~$K$ for 250~$ps$, however most surfaces were already reconstructed after 50~ps. From these 500~$ps$, we relaxed the structure with the lowest potential energy with PET as a representative of the respective Miller index.

\subsection{\label{sec:methodssoaps} Temporally Gaussian-broadened SOAP and Principal Component Analysis}
We used a Gaussian broadening from SciPy \cite{virtanenSciPy10Fundamental2020} with a standard deviation for Gaussian kernel of $\sigma$=5~\textit{ps} and 'nearest' image treatment on the edges for the SOAP vectors (cutoff=5.5~\textit{\r{A}}, Radial\textsubscript{max}=6, Angular\textsubscript{max}=8, Gaussian width = 0.3~\textit{\r{A}}, computed with featomic (available
at \\ \verb|https://github.com/metatensor/featomic|) to smoothen the noise on a MD simulation trajectory (as described above but at 500~$K$) computed every 1~$ps$.

\medskip
\textbf{Acknowledgments} \par
We thank M. L. Kellner for discussions of the surface reconstructions and the MLIP dataset.
HT gratefully acknowledges funding by the Deutsche Forschungsgemeinschaft (DFG, German Research Foundation) as a Walter-Benjamin Fellow (project number 534232519). DT acknowledges support from a Sinergia grant of the Swiss National Science Foundation (grant ID CRSII5\_202296).
MC acknowledges support by the European Research Council (ERC) under the research and innovation program (Grant Agreement No. 101001890-FIAMMA) and the NCCR MARVEL, funded by the Swiss National Science Foundation (SNSF, grant number 205602).
This work was supported by grants from the Swiss National Supercomputing Centre (CSCS) under the projects s1243, s1219 and lp26.

\medskip
\textbf{Supporting Information} \par
Supporting Information is available from the journal or from the author.

\medskip
\textbf{Conflict of Interest} \par
The authors declare no conflict of interest.

\newpage

\clearpage
\newpage
\pagebreak

\setcounter{page}{1}
\setcounter{figure}{0} 
\setcounter{table}{0} 
\setcounter{section}{0} 
\setcounter{secnumdepth}{3}

\newcommand{\beginsupplement}{%
        \setcounter{table}{0}
        \renewcommand{\thetable}{S\arabic{table}}%
        \setcounter{figure}{0}
        \renewcommand{\thefigure}{S\arabic{figure}}%
        \setcounter{section}{0}
        \renewcommand{\thesection}{S-\Alph{section}}%
        \setcounter{subsection}{0}
        \renewcommand{\thesubsection}{\arabic{subsection}}%
        \setcounter{equation}{0}
        \renewcommand{\theequation}{S-\arabic{equation}}%
     }

\beginsupplement

\onecolumngrid

{\centering
    \Large \textbf{Reconstructions and Dynamics of $\beta$-Lithium
Thiophosphate Surfaces}
\\
\centering %
Supporting Information \\
}
\newpage

\section{\label{SIsec:FF}Machine Learning Interatomic Potential}
\subsection{Lattice Parameters of $\beta$-\ch{Li3PS4}}
\FloatBarrier
 \begin{table}[!ht]
      \centering
                  \caption{Measured and computed lattice vectors of bulk $\beta$-\ch{Li3PS4}. All data is given in \textit{\AA}.}
      \begin{tabular}{ccccc}

      \hline
          Lattice Vector & PBEsol & PET & Exp. at 293~$K$\cite{mercierStructureT6trathiophosphateLithium1982} & Exp. at 637~$K$ \cite{hommaCrystalStructurePhase2011}\\
          \hline
          x & 12.87 & 12.88 & 13.066(3) & 12.8190 \\
          y &  7.81  & 7.82 & 8.015(2) & 8.2195\\
          z & 5.98 & 5.98 & 6.101(2) & 6.1236\\
          \hline
      \end{tabular}
      \label{SItab:lattice}
  \end{table}

\FloatBarrier

\subsection{Training Details and Validation}
\FloatBarrier
\begin{figure*}[!ht]
  \centering
    \includegraphics[width=0.49\linewidth]{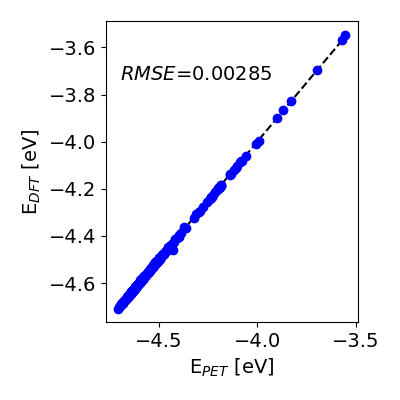}
    \includegraphics[width=0.49\linewidth]{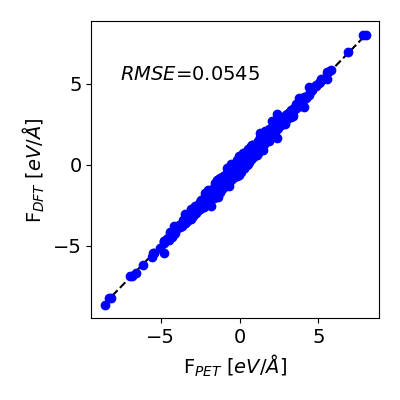}
  \caption{Energy (left) and force (right) RMSE for the PET model on a test set of LPS structures. A black dashed line with a slope of 1 is added in both plots as guide for the eye for the optimal value. }
    \label{SIfig:rmse}
\end{figure*}

\newpage
\FloatBarrier
\subsection{Validation of Simulation Results}
\FloatBarrier

\begin{figure*}[!ht]
  \centering
    \includegraphics[width=0.9\linewidth]{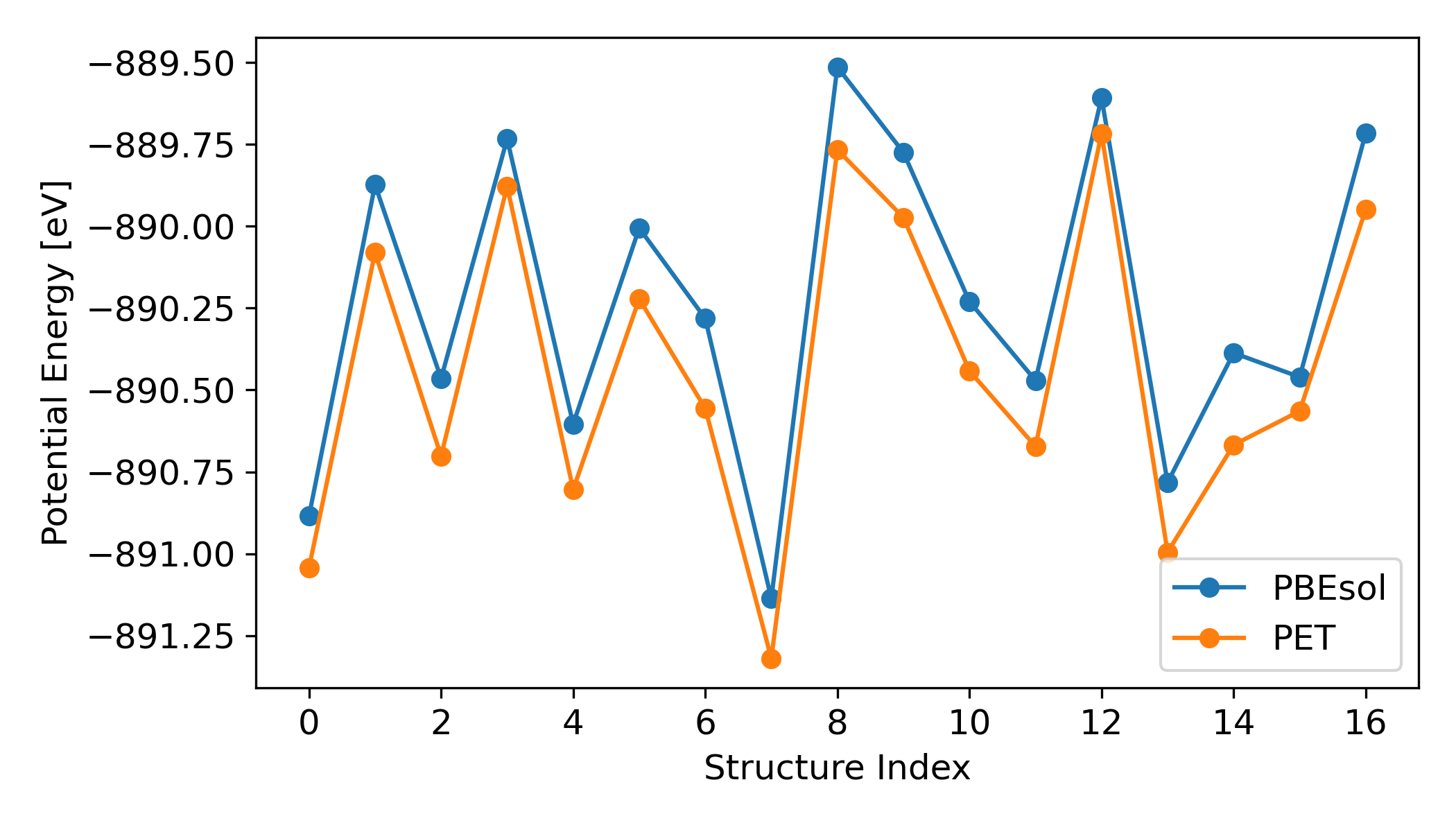}
  \caption{DFT (PBEsol) and PET energies of a (100) surface during an molecular dynamics simulation at 300~K, ambient pressure (NPT). The MD simulation was performed using PET. Then, DFT calculations were performed for the structures of the obtained trajectory.}
  \label{fig:SItetrarotation}
\end{figure*}

\begin{figure*}[!ht]
  \centering
    \includegraphics[width=0.9\linewidth]{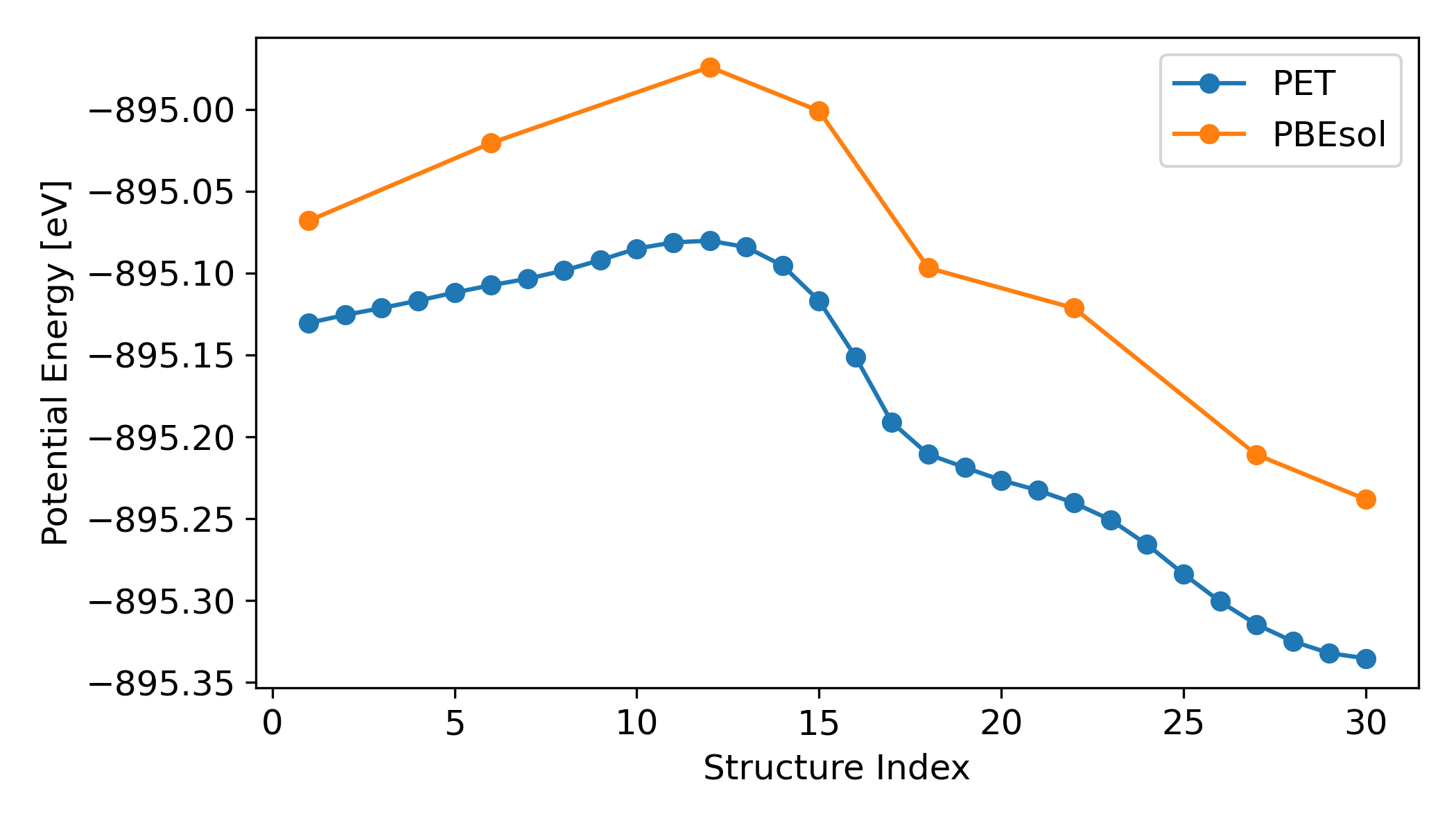}
  \caption{DFT (PBEsol) and PET energies of a Nudged-Elastic Band calculation in which a surface tetrahedron of the (100) surface rotates. The NEB calculation was performed with PET, and then single point calculations of selected structures were performed with DFT to vertify the energies.}
  \label{fig:SIflipneb}
\end{figure*}
\FloatBarrier

\clearpage
\section{\label{SIsec:functions}Computation of Surface Free Energies and Coordination Number}

We compute the surface energies $\gamma$ according to
\begin{align}
    \gamma=\frac{E_{\mathrm{surf}}-E_{\mathrm{bulk}}}{2A}
\end{align}

with the surface energy $E_{\mathrm{sufr}}$, the bulk energy $E_\mathrm{bulk}$ and the surface area~$A$.

The coordination number between atoms follows the COSINUS definition given in Plumed:
\begin{align}
s(r)=
    \begin{cases}
1&\mathrm{if}~ r\leq d_0 \\
0.5\left( cos \left(\frac{r-d_0}{r_0}\right) +1 \right) &\mathrm{if}~ d_0<r< d_0+r_0 \\
0&\mathrm{if}~ r< d_0 + r_0 
\end{cases}
\label{SIformula:coord}
\end{align}

with the $r_0$=2.0 and $d_0$=0.5.
For obtaining the sulfur ion coordination for Figure~\ref{fig:surface_stability}, the coordination of the sulfur ions in top 2~\textit{\AA} of the surface was calculated.

\FloatBarrier

  \begin{table}[!ht]
      \centering
                  \caption{Computed surface energies.}
      \begin{tabular}{cccccc}

      \hline
          Surface &Static Relax.   & MD + Relax.  & Lepley  & Marana  & Kim \\
           &S  &  & et al.\cite{lepleyStructuresLiMobilities2013} & et al. \cite{maranaComputationalCharacterizationBetaLi3PS42022} & et al.\cite{kimStructuralElectronicDescriptors2020}  \\
          &\big[$\frac{J}{m^2}$\big]&\big[$\frac{J}{m^2}$\big]&\big[$\frac{J}{m^2}$\big]&\big[$\frac{J}{m^2}$\big]&\big[$\frac{J}{m^2}$\big]\\
          \hline 
          (100) & 0.38 & 0.35   & 0.32& 0.91& 0.44  \\
          (010) & 0.59 & 0.48   &     &1.83 &0.32  \\
          (001) & 0.60 & 0.44   &     &2.15 &0.22  \\
          (110) & 0.73 & 0.42   &     &2.27 &0.28   \\
          (101) & 0.69 & 0.41   &     &8.20 &0.29  \\
          (011) & 0.42 & 0.39   &     &1.43 &0.31 \\
          (111) & 0.62 & 0.45   &     & 1.60&0.28  \\
          (210) & 0.36 & 0.36   &     & 0.99 &0.50 \\
          (021) & 0.51 & 0.46   &     &      &0.46\\
          (201) & 0.50 & 0.39   &     &      &0.47 \\
          (211) &  0.51& 0.40   &     & 1.57& \\
          \hline
      \end{tabular}
      \label{SItab:surfe}
  \end{table}

Please note that the surface energy is quite sensitive to the energy of the relaxed bulk reference structure $E_{\mathrm{bulk}}$. Here, we compare to the LPS bulk cell depicted in Figure~\ref{fig:betalps} after relaxation with the MLIP. Differences in the values can further root from the used potential to obtain the energies, where e.g. Marana et al. used PBE0 and Kim et al. used PBE.

\begin{figure*}[!ht]
  \centering
    \includegraphics[width=0.6\linewidth]{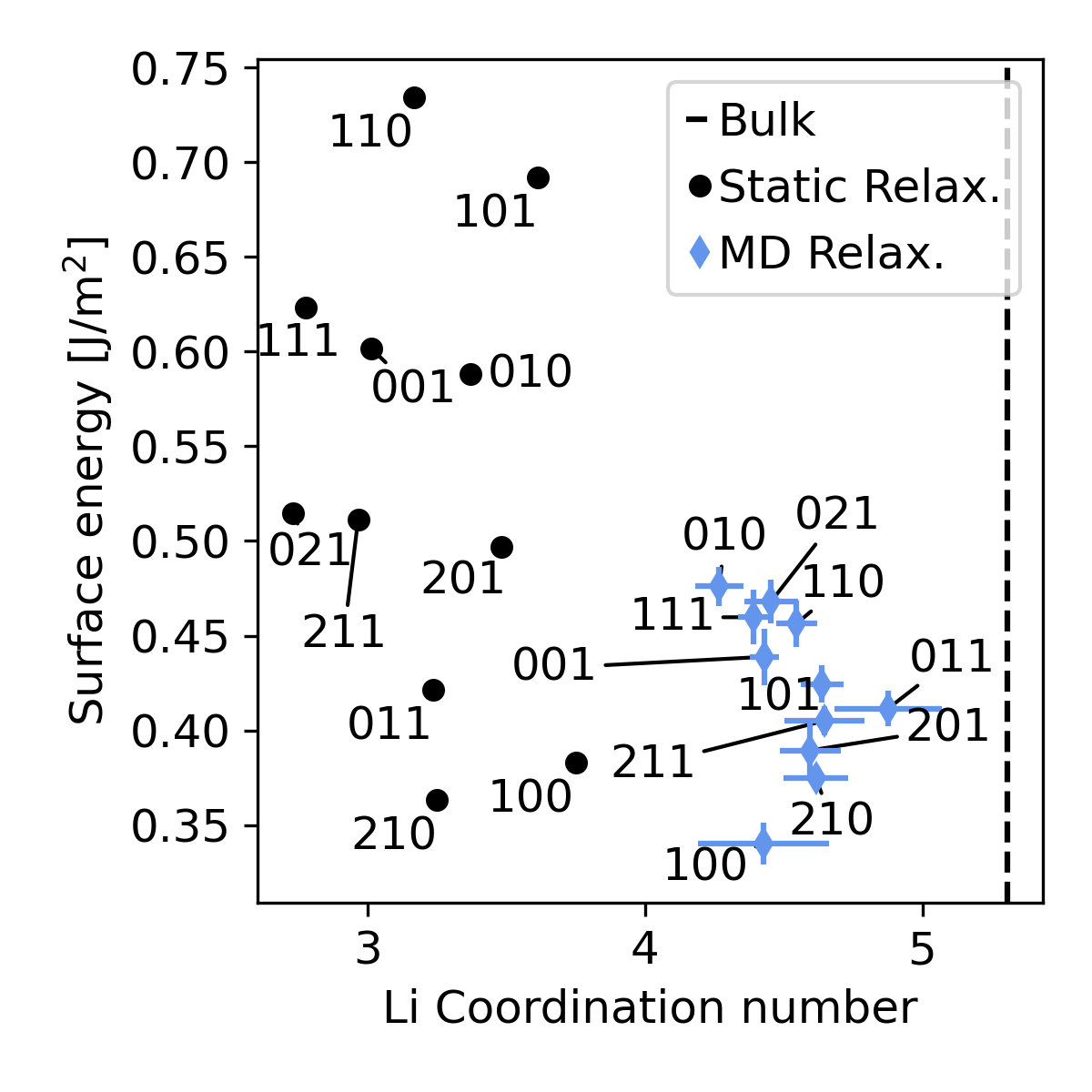}
  \caption{Surface energy of different Miller indices against the coordination of the lithium ions for the static relaxation (black dots) and MD relaxation (blue diamonds). The lithium ion coordination is defined also as explained in Section~\ref{SIsec:functions}.}
  \label{fig:SILicoord}
\end{figure*}

\newpage
\FloatBarrier

\subsection{Surface Energy of Surface Cuts with same Miller Index}
\FloatBarrier

\begin{figure*}[hbt!]
  \centering
    \includegraphics[width=0.78\linewidth]{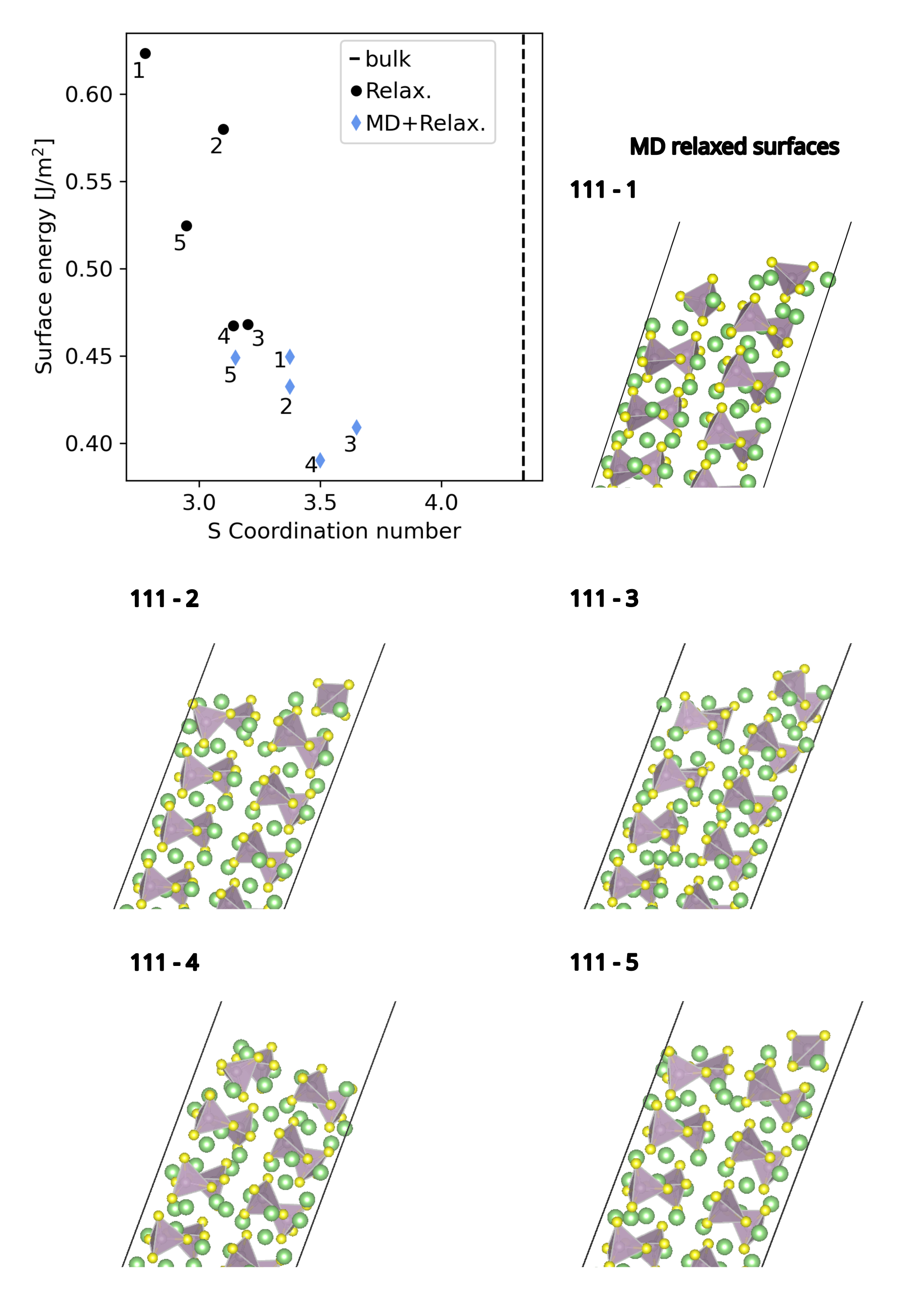}
  \caption{Surface energy of five different realizations of a (111) surface for the static relaxation (black dots) and MD relaxation (blue diamonds) (top left). The different cuts are visualized in the rest of the Figure.}
  \label{fig:SI111}
\end{figure*}

\FloatBarrier
\section{Diffusion}
\FloatBarrier

We investigated the diffusion of lithium ions at different distances to the surface at 500~$K$. The calculated bulk diffusion coefficient was 2.66~$\cdot 10^{-6}~cm^2s^{-1}$, which translates to a Nerst-Einstein ionic conductivity of 0.198 $Scm^{-1}$, compatible with the results for bulk $\beta$-LPS in the SI of Gigli \textit{et al.} \cite{gigliMechanismChargeTransport2024}.
Regarding the surface conductivity, we noted that the (010) and the (210) surface have a lower decrease in conductivity towards the surface, which could be related to less structural collapse, which is noticeable for the (210) structure in Figure~\ref{fig:surface_stability}b and little change in the surface energy.  

\begin{figure*}%
  \centering
    \includegraphics[width=0.95\linewidth]{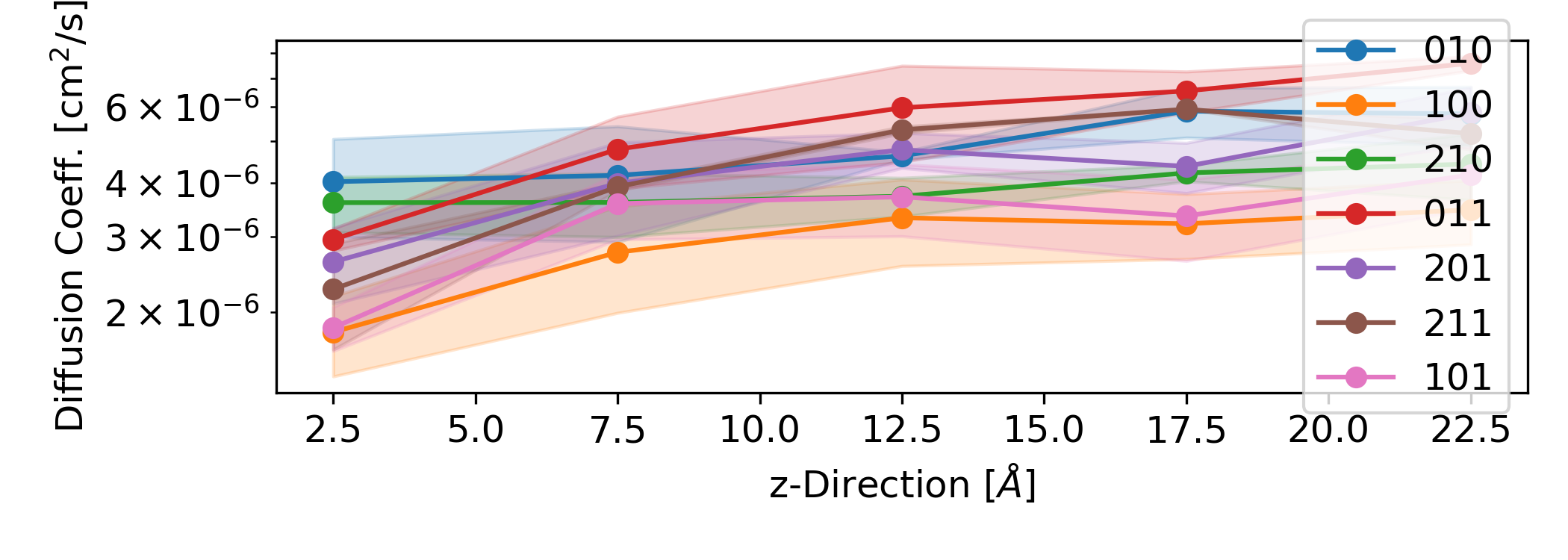}
  \caption{Diffusion coefficient of lithium ions in dependence of their distance to the surface. The standard deviation of the three MDs performed per surface is displayed as envelope.}
  \label{SIfig:diff}
\end{figure*}

\begin{figure*}%
  \centering
    \includegraphics[width=0.95\linewidth]{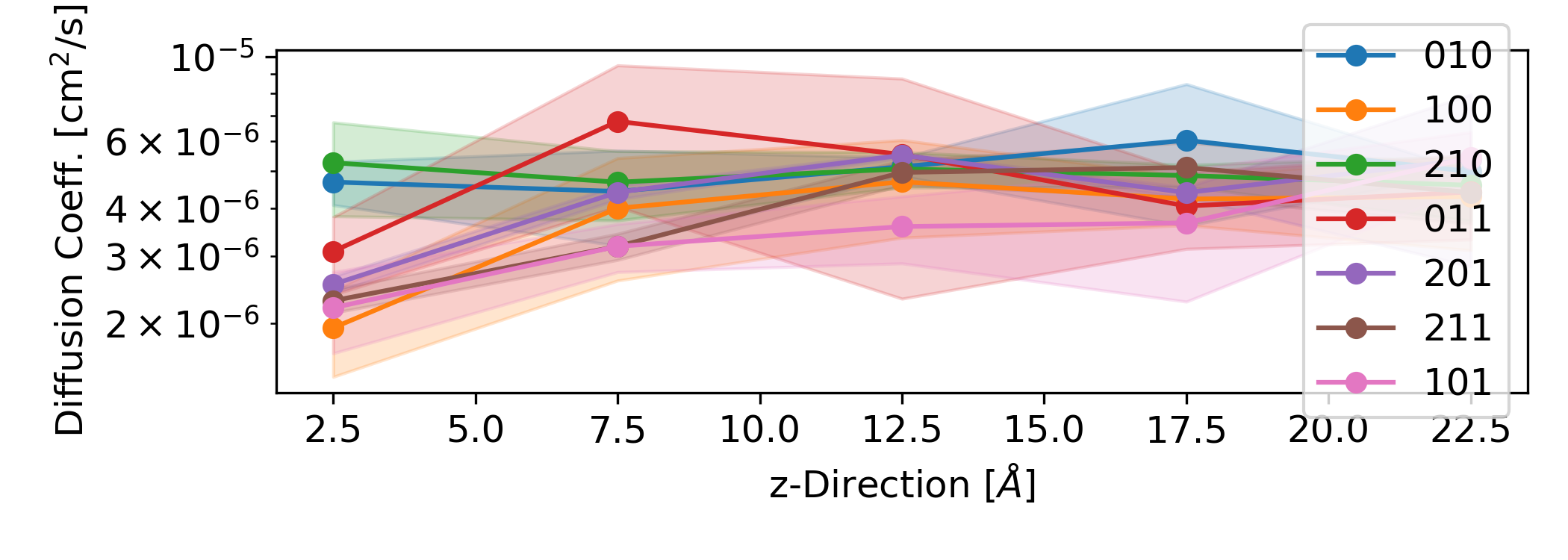} \\
    \includegraphics[width=0.95\linewidth]{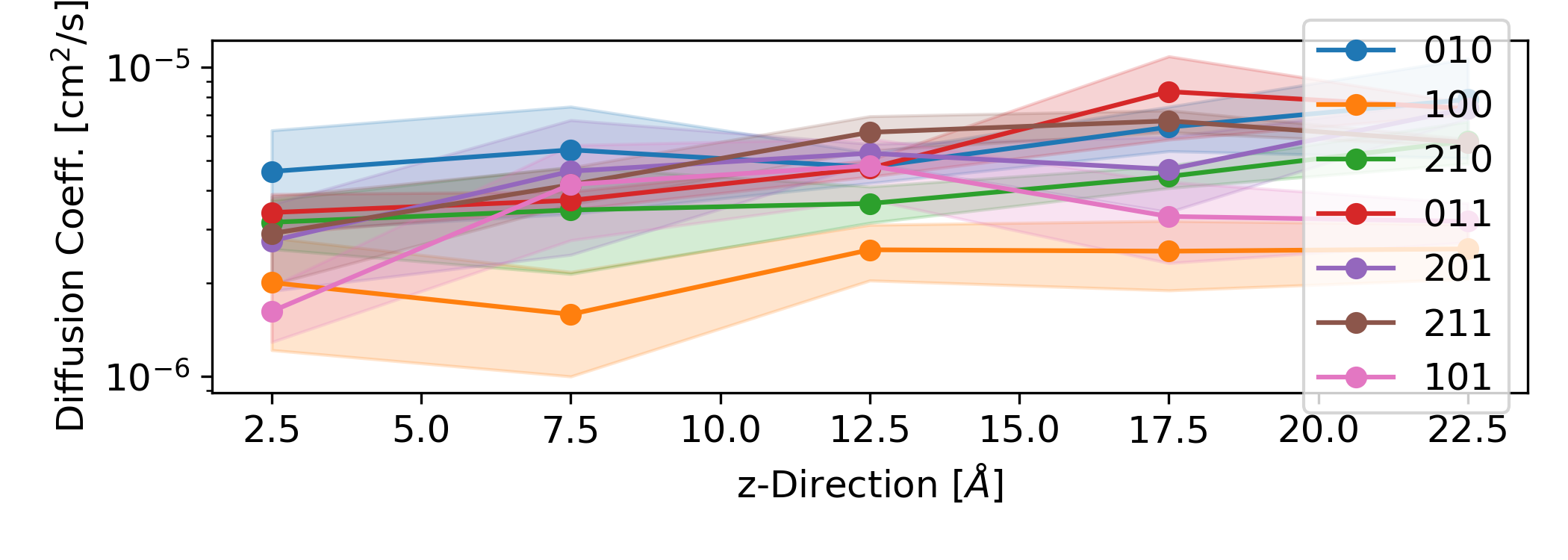} \\
    \includegraphics[width=0.95\linewidth]{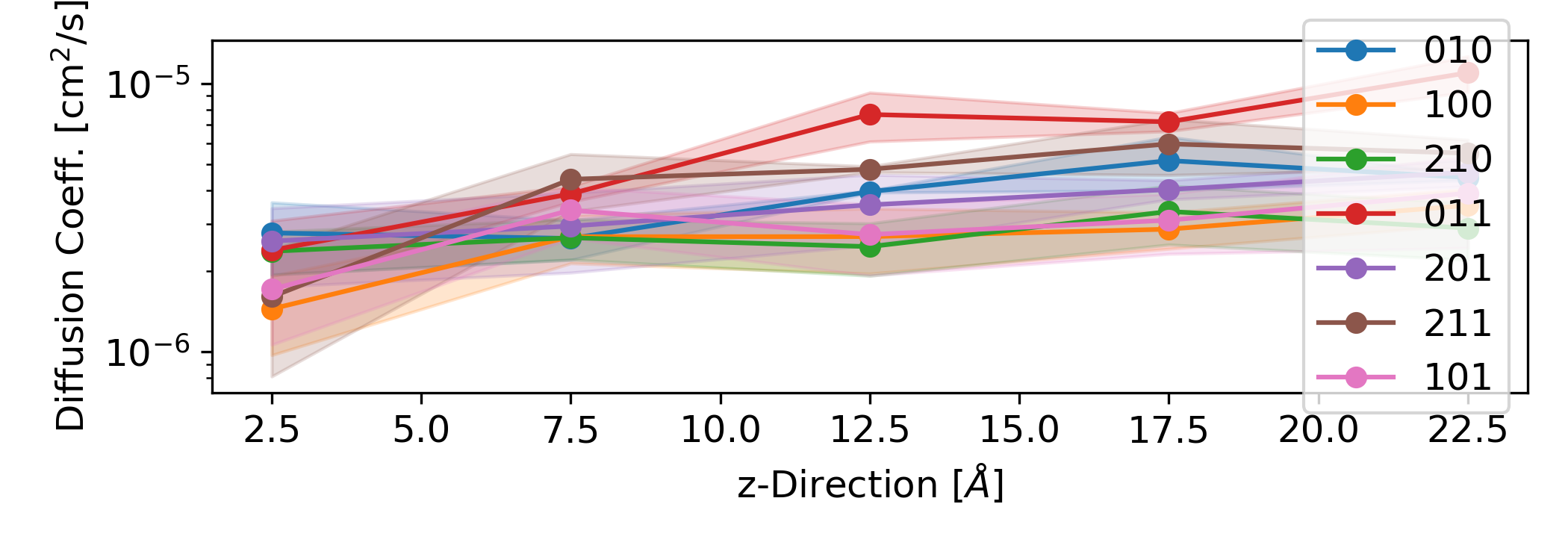} \\
    
  \caption{Direction dependent diffusion coefficient of lithium ions in dependence of their distance to the surface for diffusion in $x$-, $y$-, and $z$-direction at top, middle, and bottom, respectively. The standard deviation of the three MDs performed per surface is displayed as envelope.}
  \label{SIfig:diffxyz}
\end{figure*}
 \newpage
\FloatBarrier
\section{Wulff Constructions}
\FloatBarrier

\begin{figure*}%
  \centering
    \includegraphics[width=0.95\linewidth]{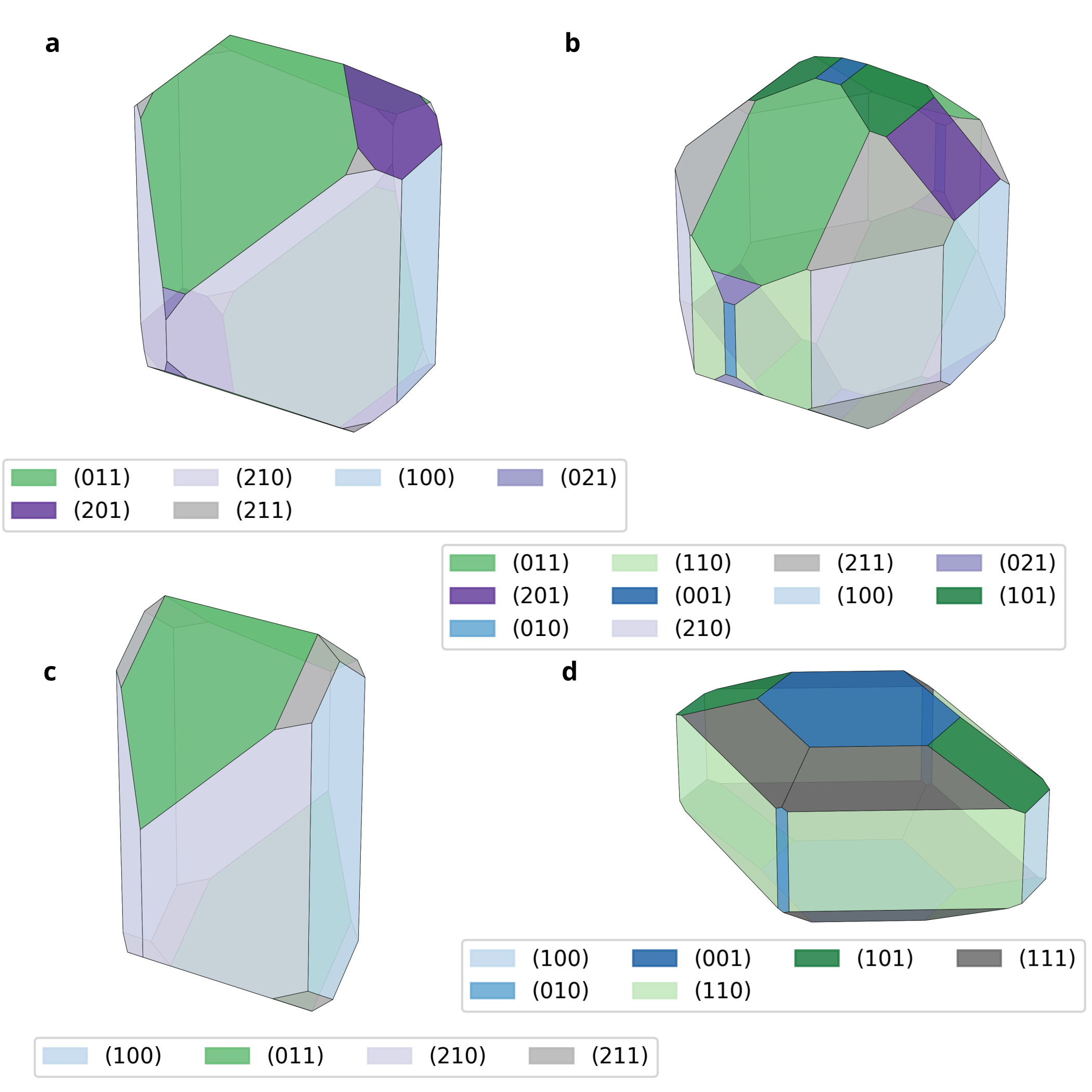}
  \caption{Wulff constructions an LPS as computed (a) in this work without MD simulation, (b) this work with MD simulation, (c) Marana et al. \cite{maranaComputationalCharacterizationBetaLi3PS42022}, and (d) according to Kim et al. \cite{kimStructuralElectronicDescriptors2020}.}
  \label{SIfig:wulff}
\end{figure*}

  \begin{table}[!ht]
      \centering
        \caption{Fractional compositions of the different surfaces for the Wulff construction shown Figure~\ref{fig:wulff} and as computed from the free surface energies provided by Marana et al. \cite{maranaComputationalCharacterizationBetaLi3PS42022} and Kim et al.\cite{kimStructuralElectronicDescriptors2020}.}
      \begin{tabular}{crrrr}
      \hline
          Surface & Static Relax. & MD Relax. & Marana et al. \cite{maranaComputationalCharacterizationBetaLi3PS42022} & Kim et al.\cite{kimStructuralElectronicDescriptors2020} \\
          &[\%]&[\%]&[\%]&[\%]\\
          \hline
          (100)&9.45&  11.82  & 19.68 & 3.83  \\
          (010)&  - & 0.47  &   -   & 1.06  \\
          (001)&  - &0.70   &    -  & 17.34  \\
          (110)&  - & 8.61  &    -  & 36.33  \\
          (101)&  - &   5.75   &    -  & 10.68  \\
          (011)&36.91& 23.34  & 30.79 & -   \\
          (111)&   - & -   &  - & 30.76 \\
          (210)&43.48& 19.53  & 44.05 & -   \\
          (021)&0.33& 0.80    &   -   & -   \\
          (201)&9.12& 9.42   &   -   & -   \\
          (211)& 0.65&19.57 &5.48 & \\
          \hline
      \end{tabular}
      \label{SItab:wulff}
  \end{table}

\FloatBarrier

\section{Surface State analysis}

\subsection{Principal component analysis}
\FloatBarrier

\begin{figure*}[!ht]
  \centering
    \includegraphics[width=1\linewidth]{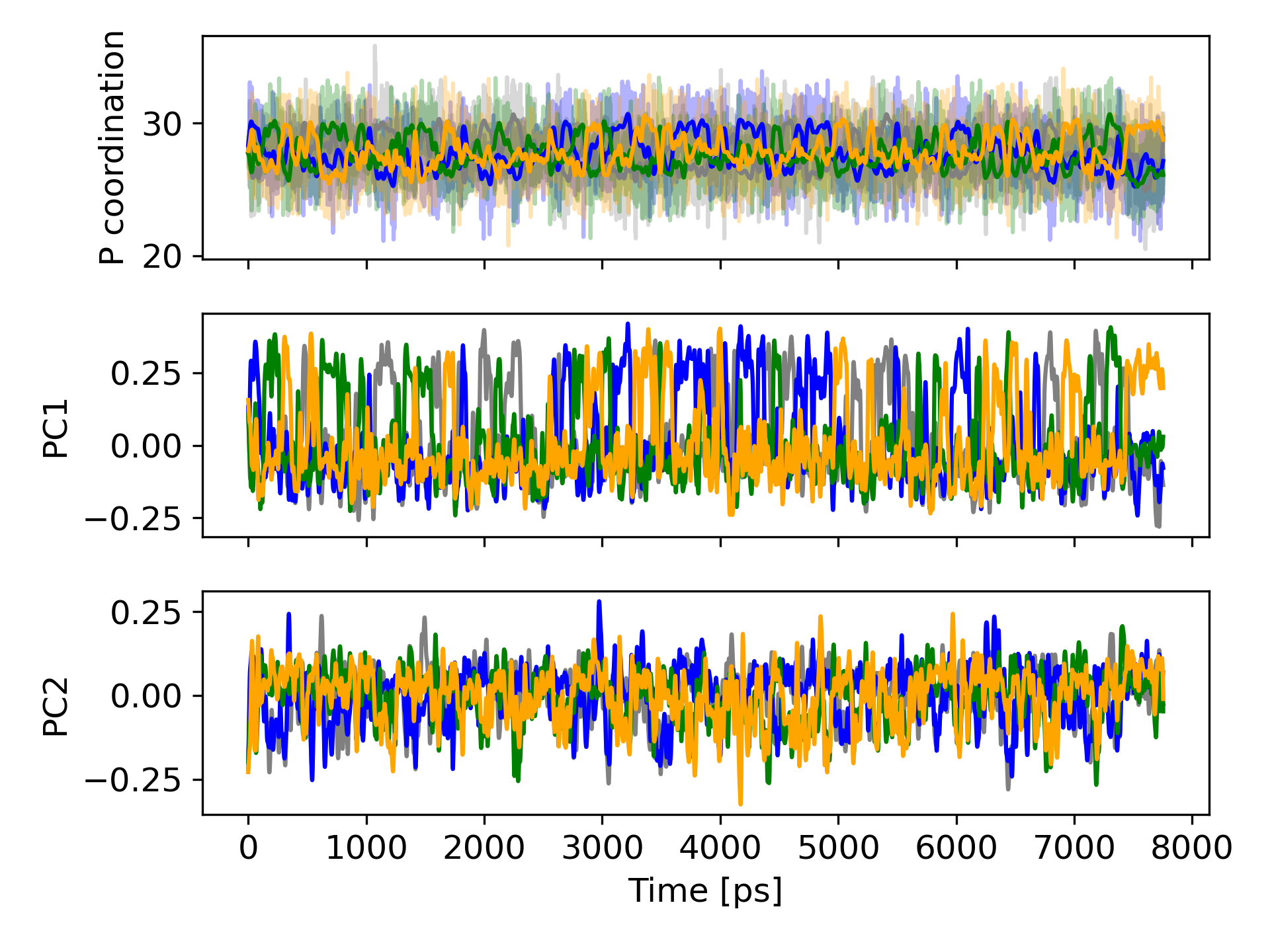}
  \caption{Temporal evolution of the different principal components (PCs) of the phosphorous atomic environments and the total coordination number for the small (100) surface shown in Figure~\ref{fig:soaps}. The solid lines are temporally broadened, and for the coodination of the phosphorous the original data is slightly more transparent in the background.  Different colors belong to the different tetrahedra on the surface (compare Figure~\ref{fig:soaps}).}
  \label{fig:SIpc_coordination}
\end{figure*}

\begin{figure*}[!ht]
  \centering
    \includegraphics[width=0.49\linewidth]{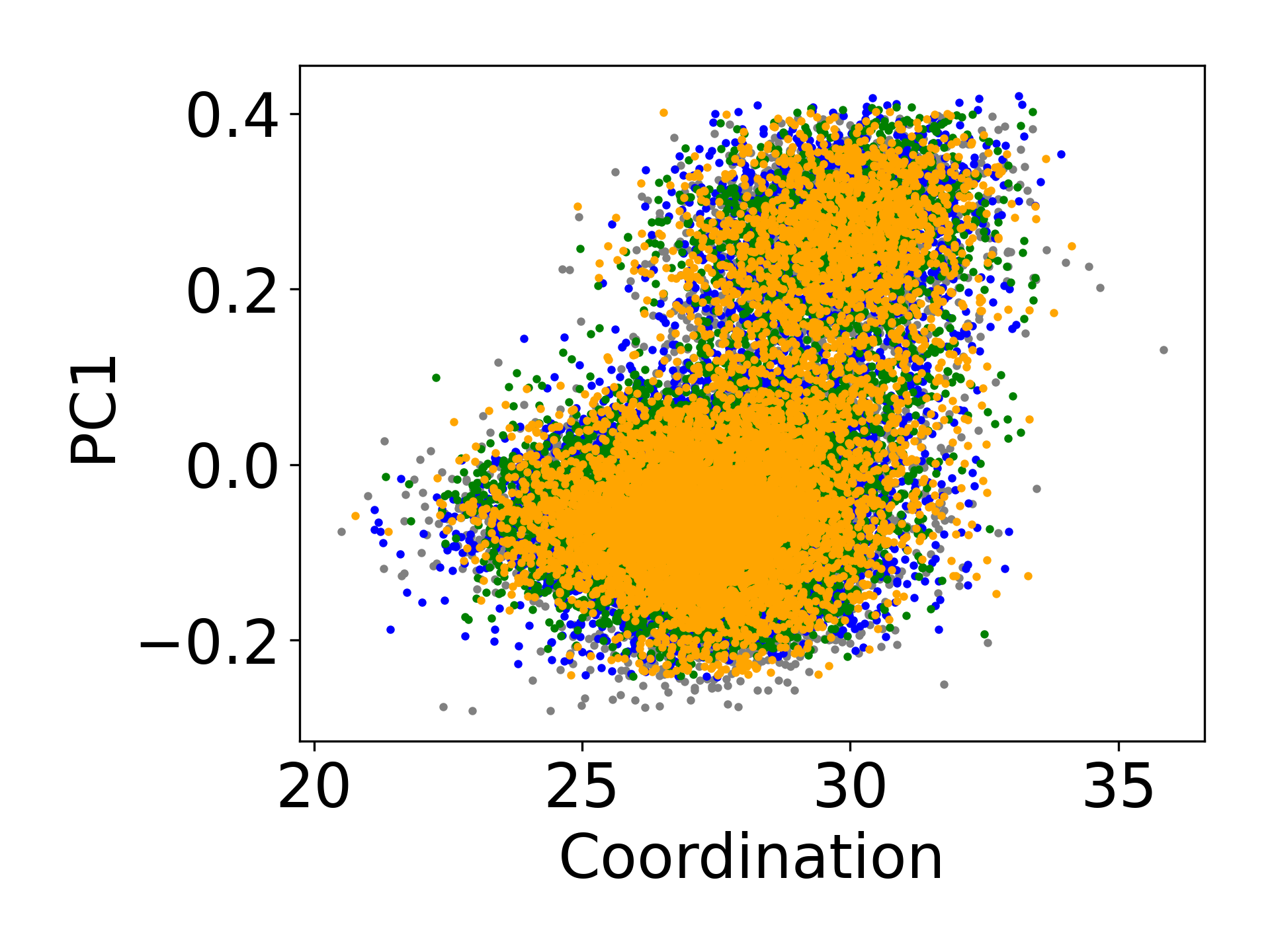}
    \includegraphics[width=0.49\linewidth]{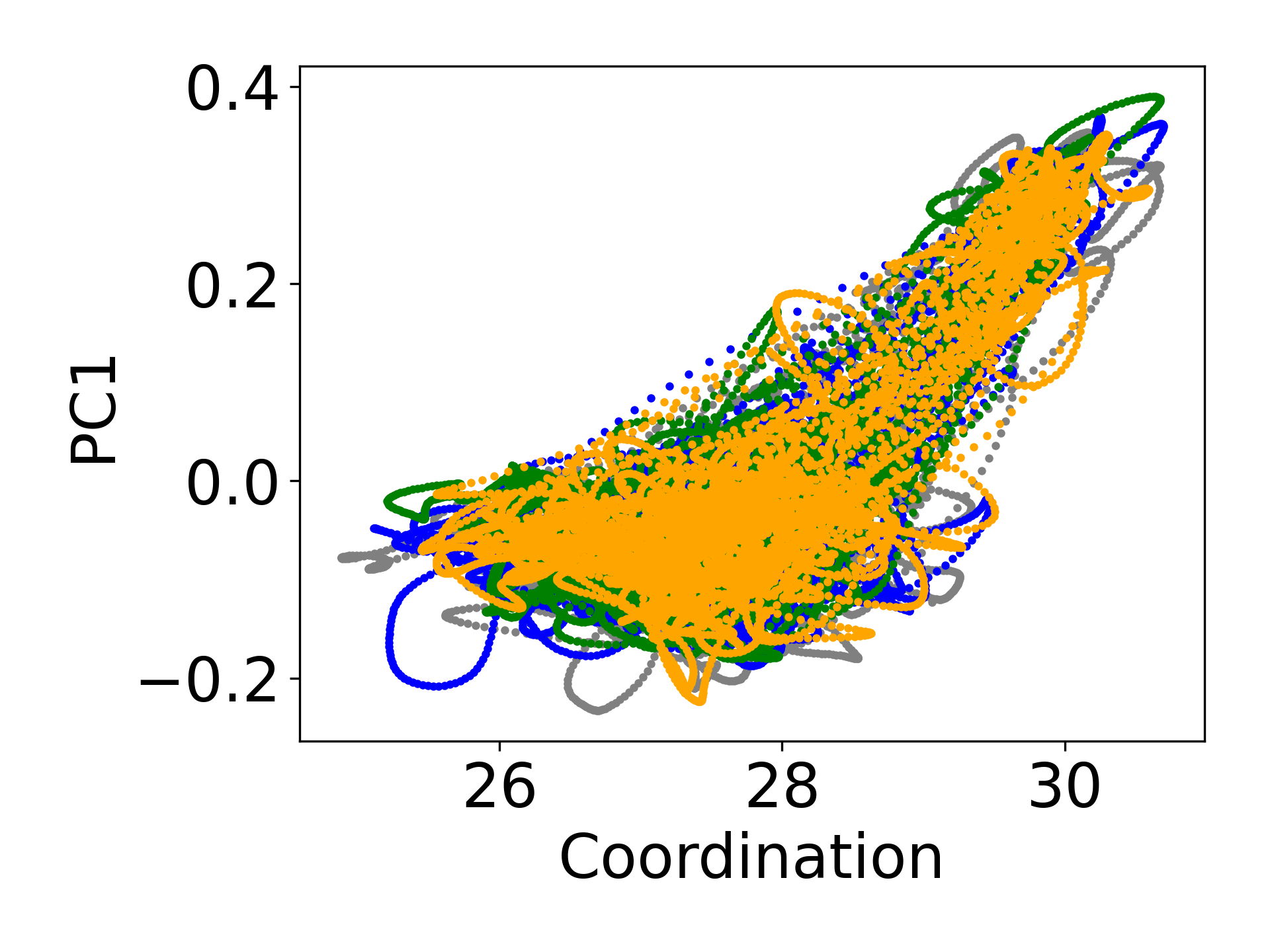}
  \caption{Parity plot of the SOAP descriptors of the surface tetrahedra without (left) and with (right) temporal averaging as used in Figure~\ref{fig:soaps} against the coordination with all other elements of the P centered in the respective tetrahedron. The coordination was computed according to Formula~\ref{SIformula:coord} with a cutoff of 5.5~\textit{\AA}.}
  \label{fig:SIpc1coord}
\end{figure*}

\FloatBarrier
\subsection{Finite Size Effects \label{SIsec:finitesize}}
\FloatBarrier
The correlating movement of neighboring tetrahedra led us to want to explore whether we have finite size effects in the computed trajectory. For this reason, we ran molecular dynamics simulations at 500~$K$ of 2$\times$2$\times$1 supercells, one of the initial clean cut of the surface and one as a replicate of the smaller (100) surface with one tetrahedron already flipped.

\begin{figure*}[!ht]
  \centering
    \includegraphics[width=1\linewidth]{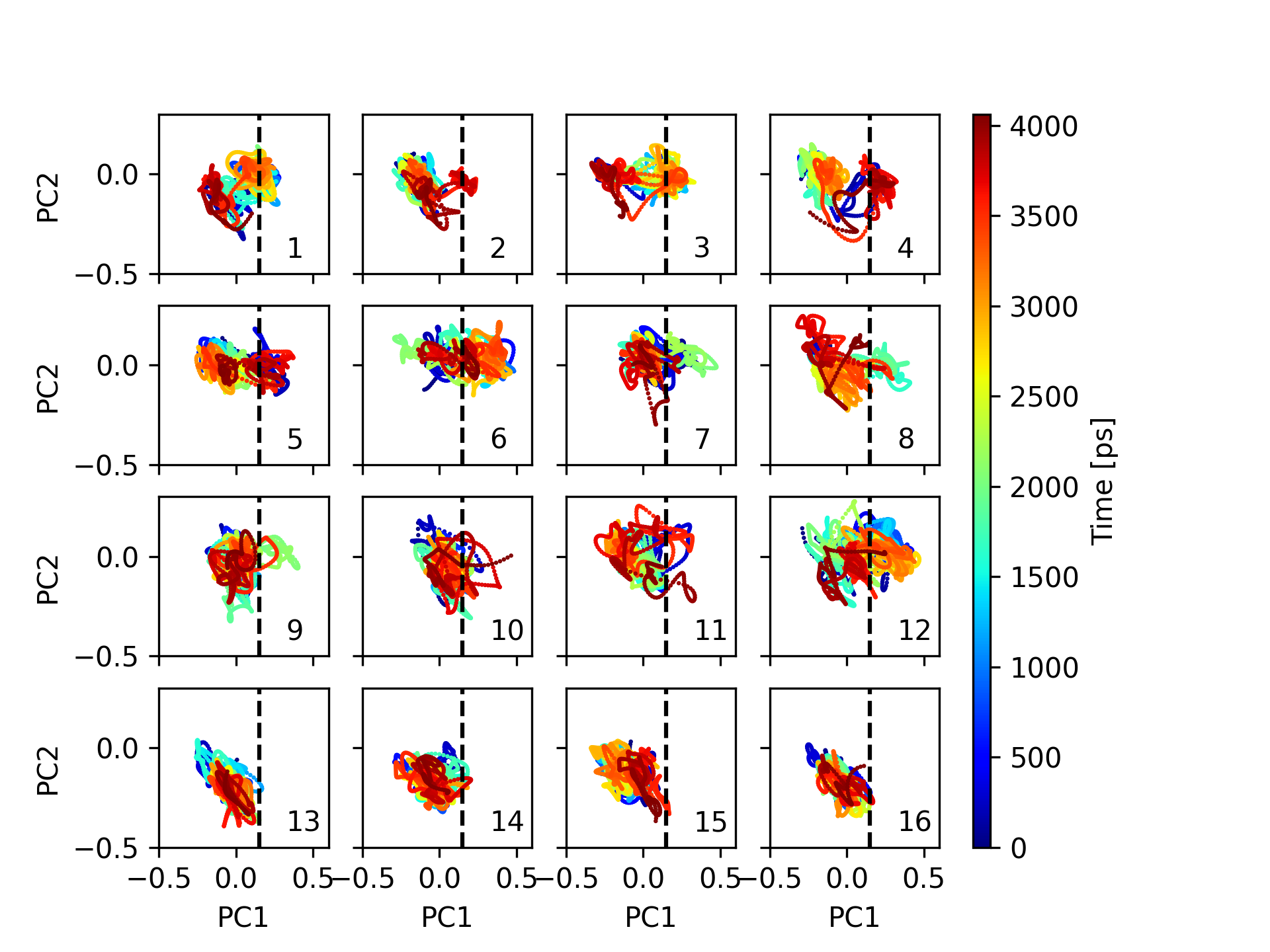}
  \caption{Temporally averaged SOAP vectors of a 500~$K$ MD simulation
  transformed according to the principle component analysis performed for the small (100) cell (compare Figure~\ref{fig:soaps}) for the (100) surface of an initially clean surface cut.}
  \label{fig:SIsoaps_large_cleancut}
\end{figure*}

\begin{figure*}[!ht]
  \centering
    \includegraphics[width=1\linewidth]{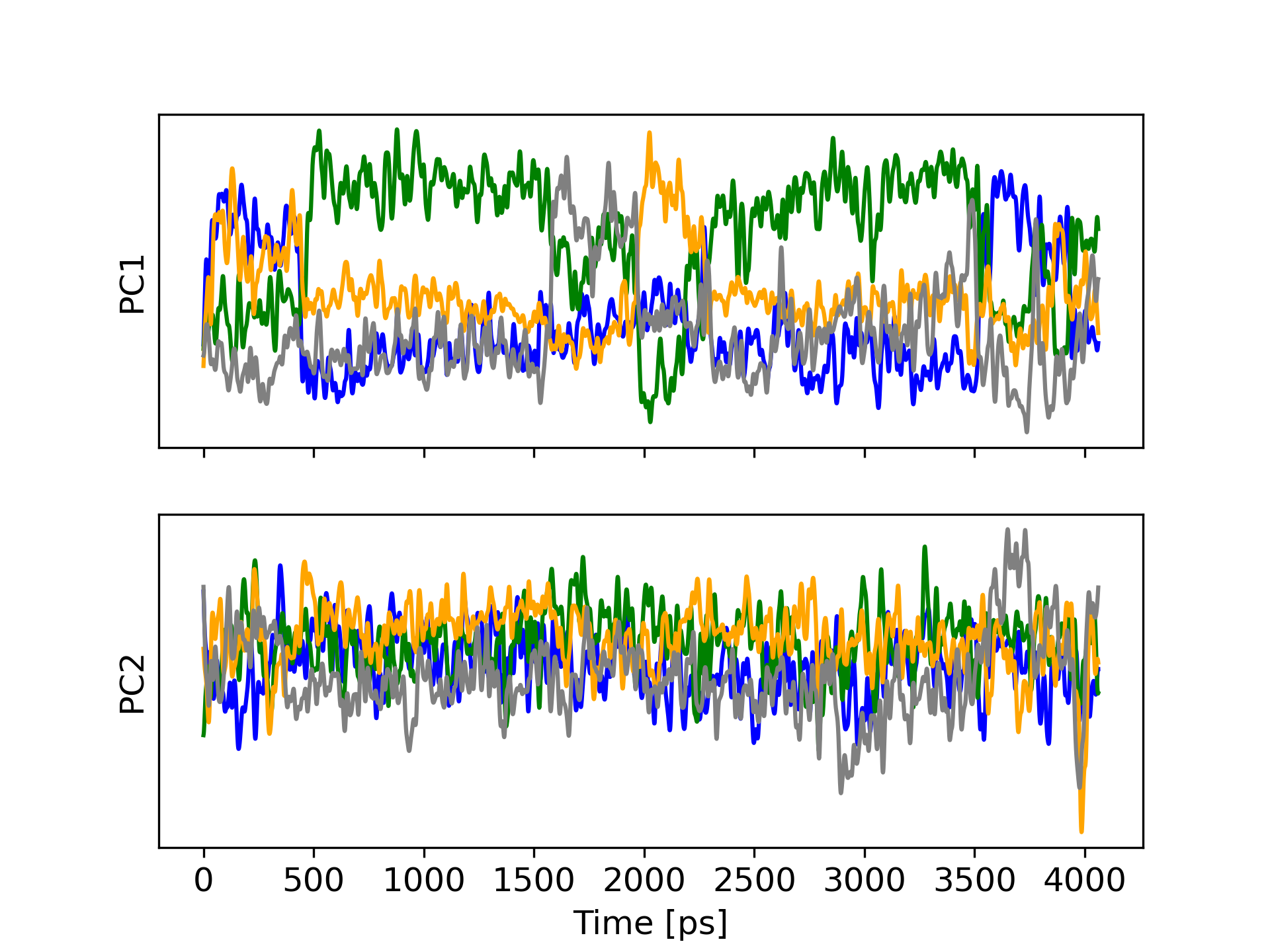}
  \caption{Temporal evolution of the PC1 and PC2 component of tetrahedron 5-8 of Figure~\ref{fig:SIsoaps_large_cleancut}. Colors for the individual surface tetrahedra are according to Figure~\ref{fig:soaps}.}
  \label{fig:SIpc_cleancut}
\end{figure*}

\begin{figure*}[!ht]
  \centering
    \includegraphics[width=1\linewidth]{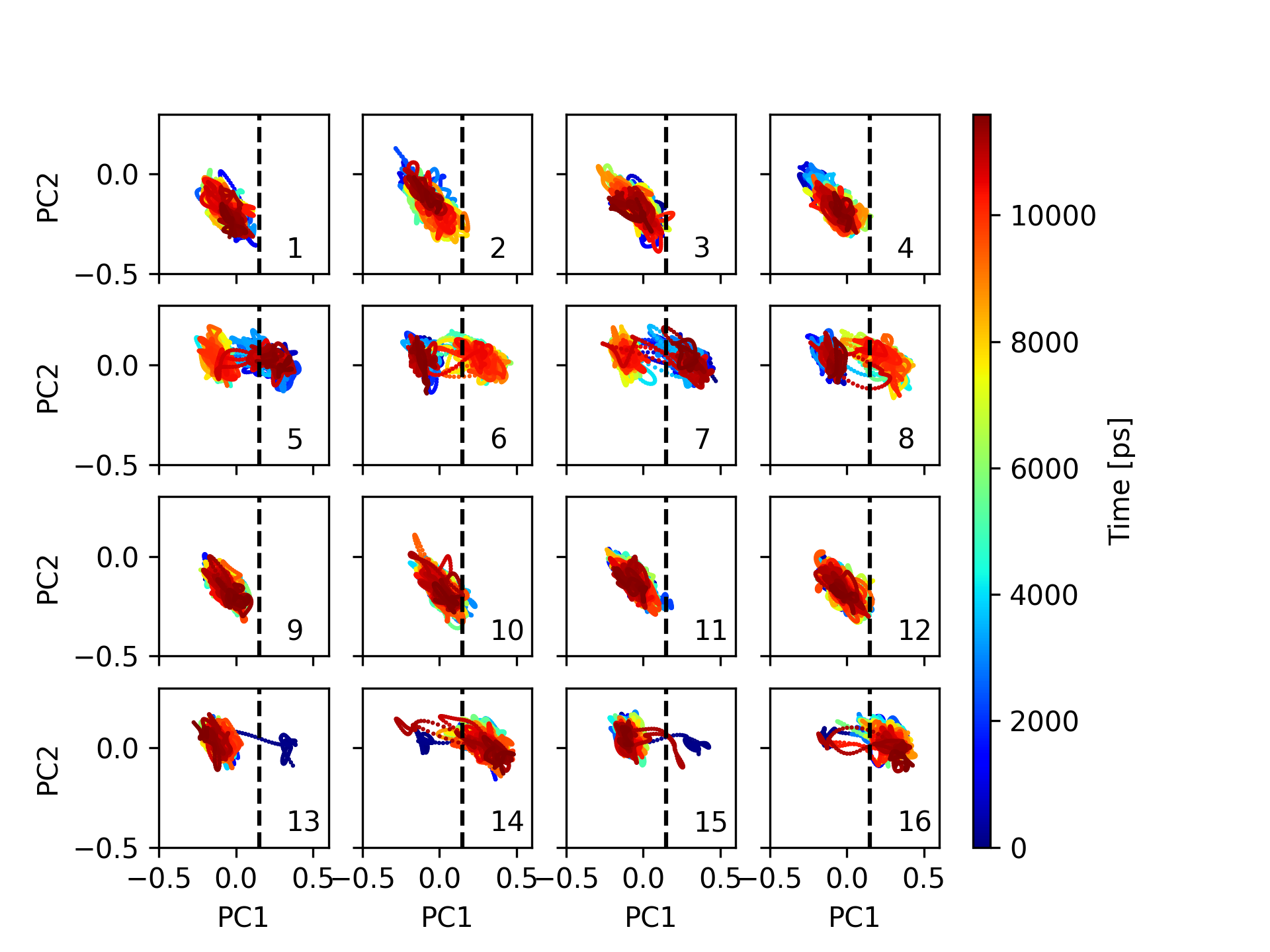}
  \caption{Temporally averaged SOAP vectors of a 500~$K$ MD simulation
  transformed according to the principle component analysis performed for the small (100) cell (compare Figure~\ref{fig:soaps}) for the (100) surface of the 2$\times$2$\times$1 supercell of the small (100) surface with one tetrahedron flipped to that its edge is exposed to the surface.}
  \label{fig:SIsoaps_large}
\end{figure*}

\begin{figure*}[!ht]
  \centering
    \includegraphics[width=1\linewidth]{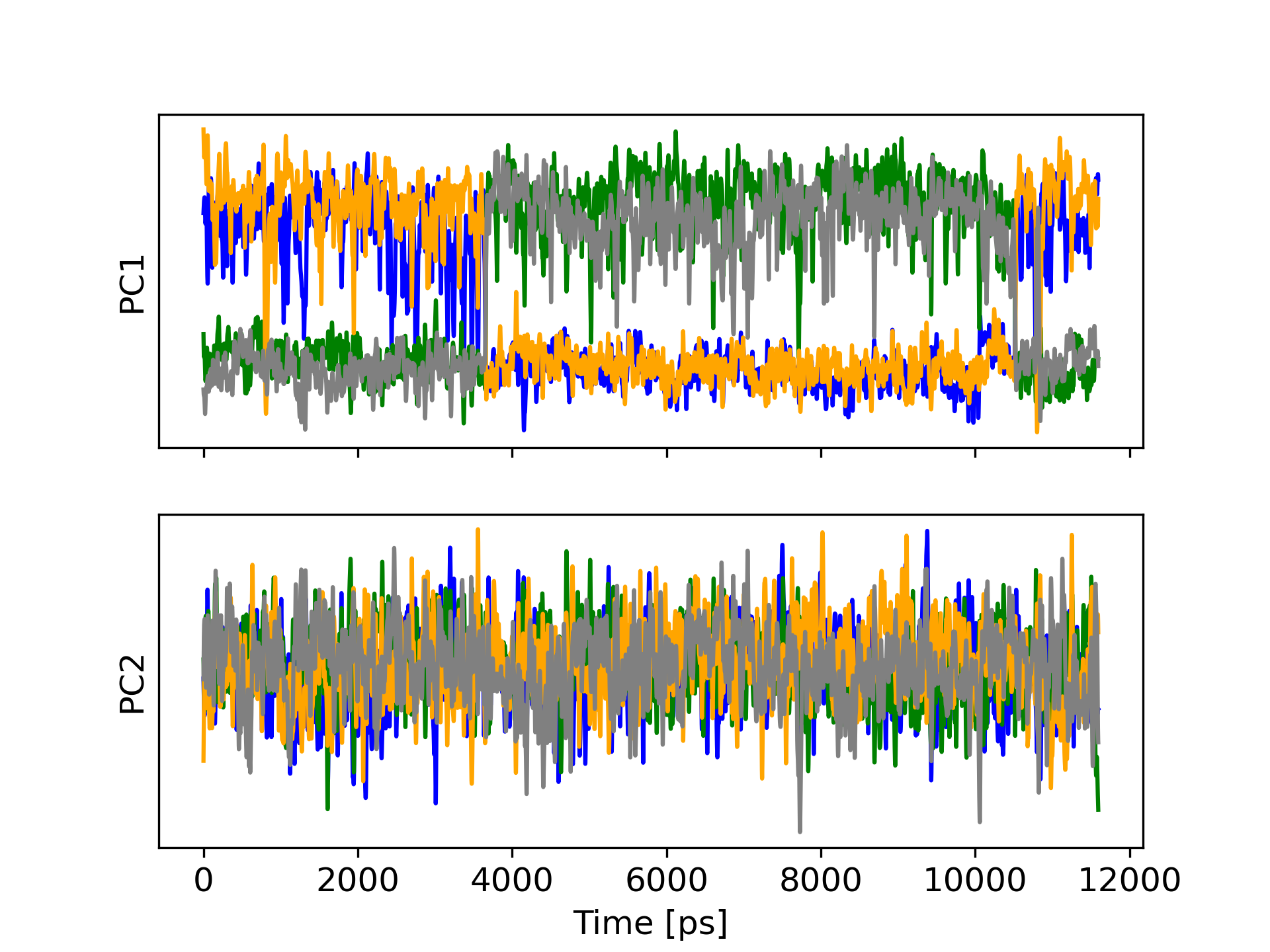}
  \caption{Temporal evolution of the PC1 and PC2 component of tetrahedron 5-8 of Figure~\ref{fig:SIsoaps_large}. Colors for the individual surface tetrahedra are according to Figure~\ref{fig:soaps}.}
  \label{fig:SIpc_large}
\end{figure*}

We observed rotation of the surface tetrahedra in both enlarged cells (compare Figures~\ref{fig:SIsoaps_large_cleancut} and \ref{fig:SIsoaps_large} for the temporal evolution of the individual tetrahedra, and Figure~\ref{fig:SIpc_cleancut} and \ref{fig:SIpc_large}, respectively, for the evolution of PC1 and PC2); however the flipping frequency was lower, especially for the already reconstructed surface. This leads us to speculate that there might be some more or less stable surface patterns, however, even these are still observed to be dynamic on relevant time scales (which we assess in the following section).

\FloatBarrier
\subsection{Calculation of Surface Dynamics \label{SIsec:dynamics}}
\FloatBarrier
The MD shown in Figure~\ref{fig:soaps} was conducted at 500~$K$ to efficiently sample the transition. In order obtain the rate constant $k$ of the rotation of the surface tetrahedra at a different temperature $T$, we can use the Arrhenius equation  

\begin{equation}
    \frac{k_1}{k_2}=\frac{Ae^{-\frac{E_{\mathrm A}}{k_B T_1}}}{Ae^{-\frac{E_{\mathrm A}}{k_B T_2}}}
\end{equation}
and solve it for the rate constant $k_1$ of interest
\begin{equation}
    k_1=k_2\frac{E_{\mathrm A}}{k_B (T_1-T_2)}  .
\end{equation}

This includes the Boltzmann factor $k_B$, the activation energy $E_{\mathrm{A}}$,
and the frequency factor $A$, whose temperature dependence is negligible here in comparison to the exponential term.
The rate constant $k_2$, which corresponds to the one at 500~$K$, can be obtained from the simulations, which we explain in the following. As PC1 allows an easy differentiation of the surface orientations of the tetrahedra (compare Figure~\ref{fig:SIpc_coordination}), one can easily count the transitions for the given time.  
Here, we want to do a rather conservative estimation of the surface dynamics, to see if the surface is still dynamic on relevant time scales. Thus we counted the tetrahedra flips from the 4 tetrahedra displayed in Figure~\ref{fig:SIpc_large} and found at least 12 crossings between the surface states for the 4 tetrahedra over the simulation time of almost 12~$ns$. The activation energy $E_{\mathrm A}$ was estimated as upper bound from the nudge elastic band (NEB) calculations presented in Figure~\ref{fig:SIflipneb}.

\begin{equation}
    k_1=\frac{12}{12 \dot 4} ns^{-1} \frac{0.3~eV}{k_B (300-500)}=6.89 s^{-1}
\end{equation}
The reaction rate $k_1$ allows us to compute also the lifetime $\tau$ of the state at 300~$K$:
\begin{equation}
    \tau=1/k_1=1/6.89 s^{-1} = 145~ms  .
\end{equation}

\begin{figure*}
  \centering
    \includegraphics[width=0.45\linewidth]{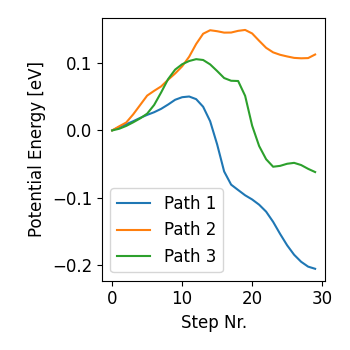}
        \includegraphics[width=0.45\linewidth]{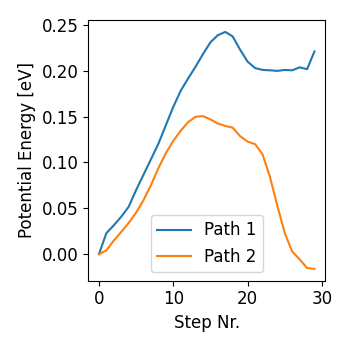}
  \caption{Nudged elastic band calculations of one tetrahedron rotating on the surface (left) and two neighbored tetrahedra rotating on the surface (right) of a small (100) surface like in Figure~\ref{fig:soaps}. In the case of the single rotation, step 0 is always belonging to a state where the tetrahedron exposes a plane, and step 30 to where it exposes an edge. For all surface rotations, the barrier is between 100-250~$meV$. Path~1 fot the single flip is the same as in Figure~\ref{fig:SIflipneb}.}
  \label{fig:SIcomparisonflips}
\end{figure*}

\begin{figure*}[!ht]
  \centering
    \includegraphics[width=\linewidth]{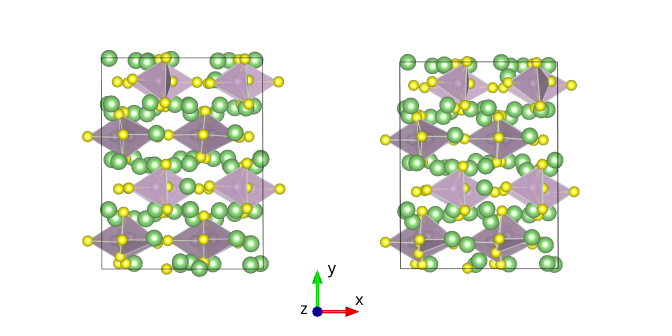}
  \caption{Initial (left) and final (right) state of the Path~3 in Figure~\ref{fig:SIcomparisonflips} (left). The tetrahedron on the top right rotates from an exposed plane to an exposed edge.}
  \label{fig:SInebsurfaces}
\end{figure*}

\begin{figure*}[ht]
  \centering
    \includegraphics[width=0.45\linewidth]{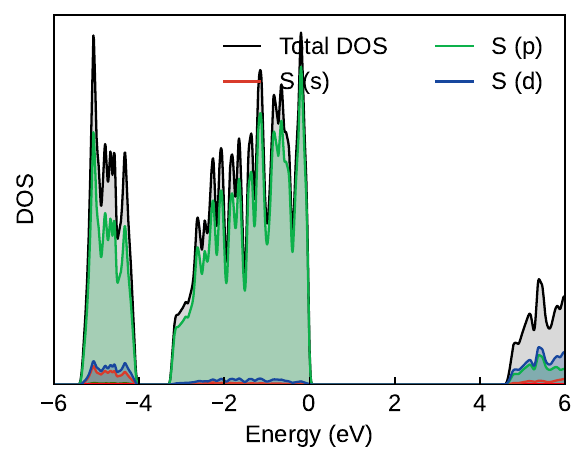} \\
     \includegraphics[width=0.45\linewidth]{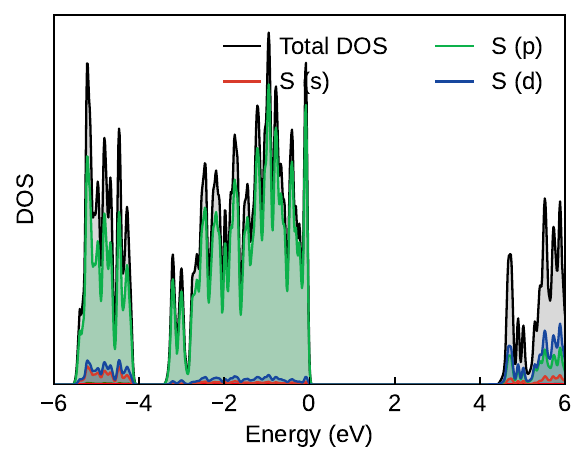}
     \includegraphics[width=0.45\linewidth]{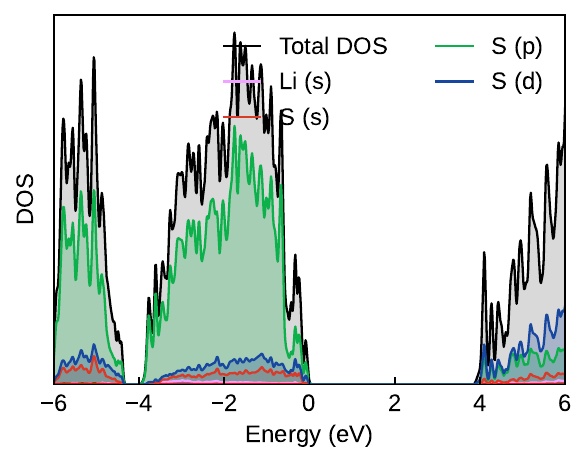}
  \caption{PDOS of bulk LPS (top), the static relaxed LPS (100) surface (bottom left) and the MD relaxed surface (bottom right). Visualized with sumo \cite{mganoseSumoCommandlineTools2018}.}
  \label{fig:SIpdos_bulk}
\end{figure*}

The NEB calculations also reflect the complexity to assess the surface states quite well, as the energy barrier of the rotation as well as its final state differ largely for the individual calculations. The initial and final structures for these calculations were all obtained from MD snapshots followed by relaxation of the structure. Often we found here that initial and final path converged to the same structure, indicating a non-existent or very low barrier between these states. This volatility is most likely also related to the redistribution of the lithium ions when the surface tetraheda, as they are found on distinct positions around the unrotated/rotated tetrahedra.
As an example, the initial and final state of Path~2 of the NEB calculation shown in Figure~\ref{fig:SIcomparisonflips} (left) are shown in Figure~\ref{fig:SInebsurfaces}.

\newpage
\FloatBarrier
\section{Electronic Structure of Surface}
\FloatBarrier

\begin{figure*}[ht]
  \centering
    \includegraphics[width=0.45\linewidth]{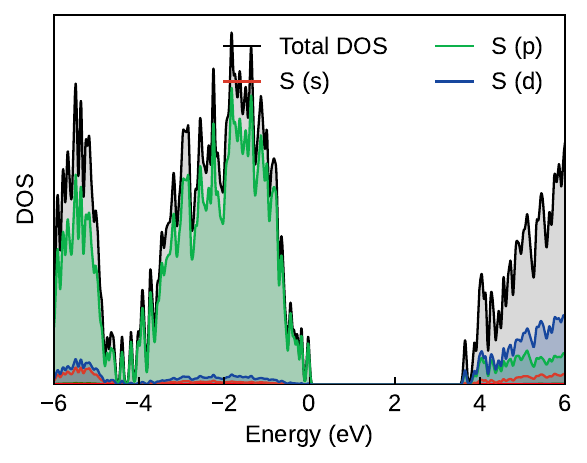} 
     \includegraphics[width=0.45\linewidth]{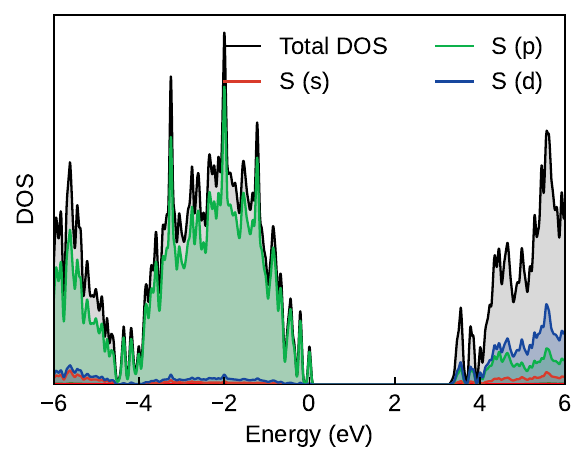} \\
     \includegraphics[width=0.45\linewidth]{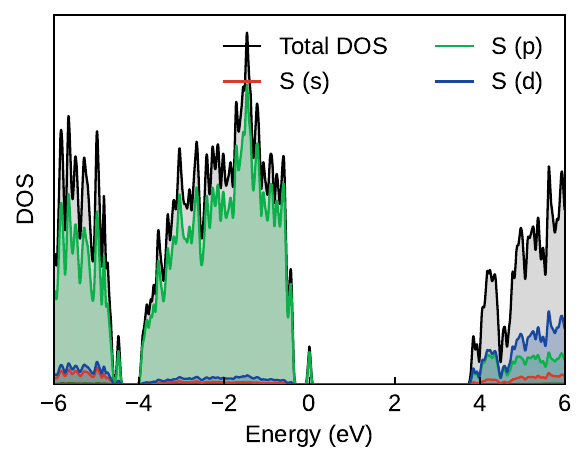}
    \includegraphics[width=0.45\linewidth]{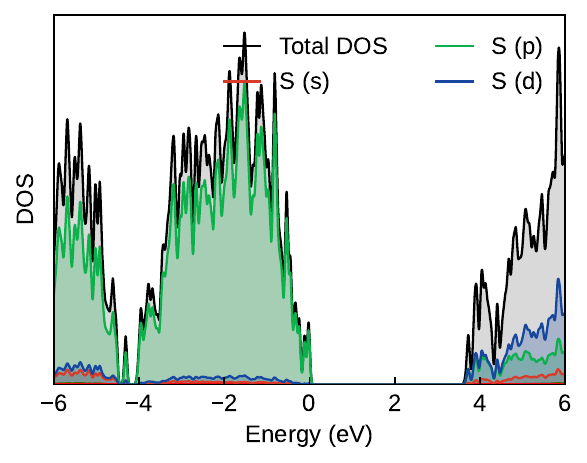}
  \caption{pDOS of MD snapshots after 50~$ps$ (top left), 1000~$ps$ (top right), 2000~$ps$ (bottom left), and 7000~$ps$ (bottom right). These DOS were averaged to obtain the DOS displayed in Figure~\ref{fig:dos}a. Visualized with sumo \cite{mganoseSumoCommandlineTools2018}.}
  \label{fig:SIpdos_md}
\end{figure*}

\begin{figure*}[ht]
  \centering
    \includegraphics[width=0.7\linewidth]{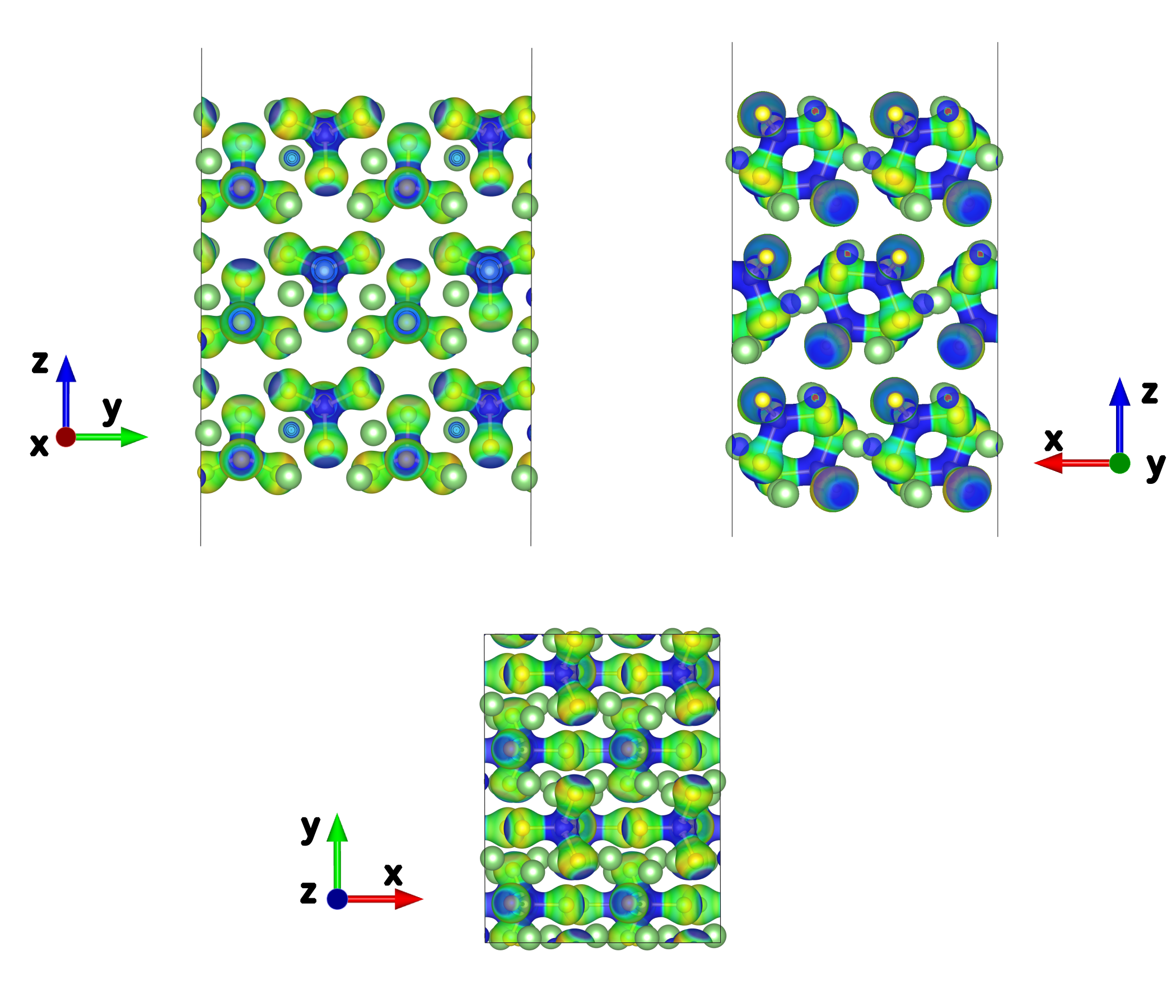} \\
  \caption{Electrostatic potential of the statically relaxed structure.}
  \label{fig:SIsimple}
\end{figure*}

\begin{figure*}[ht]
  \centering
    \includegraphics[width=0.7\linewidth]{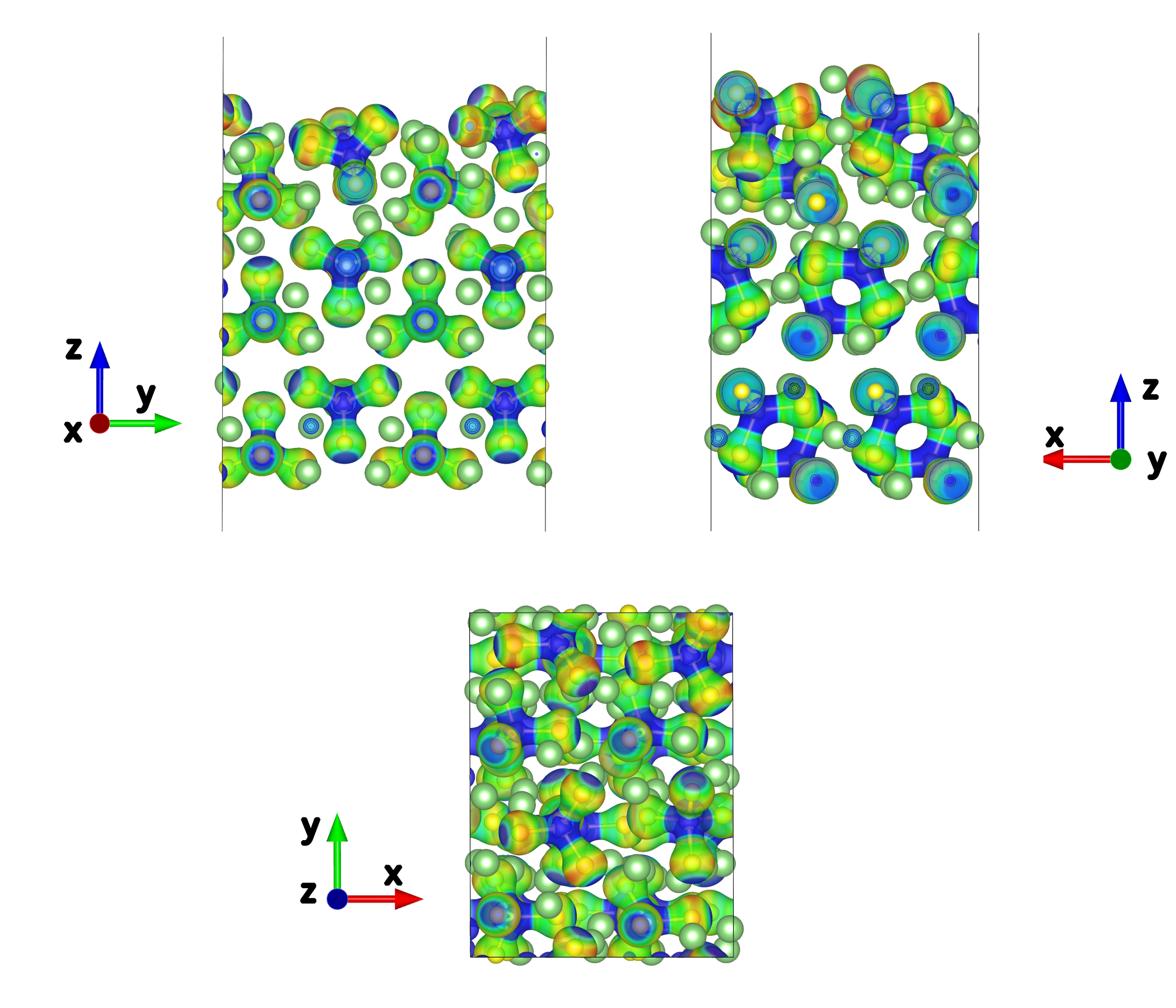} \\
  \caption{Electrostatic potential of the MD relaxed structure.}
  \label{fig:SImdrelax}
\end{figure*}

\begin{figure*}[ht]
  \centering
    \includegraphics[width=0.7\linewidth]{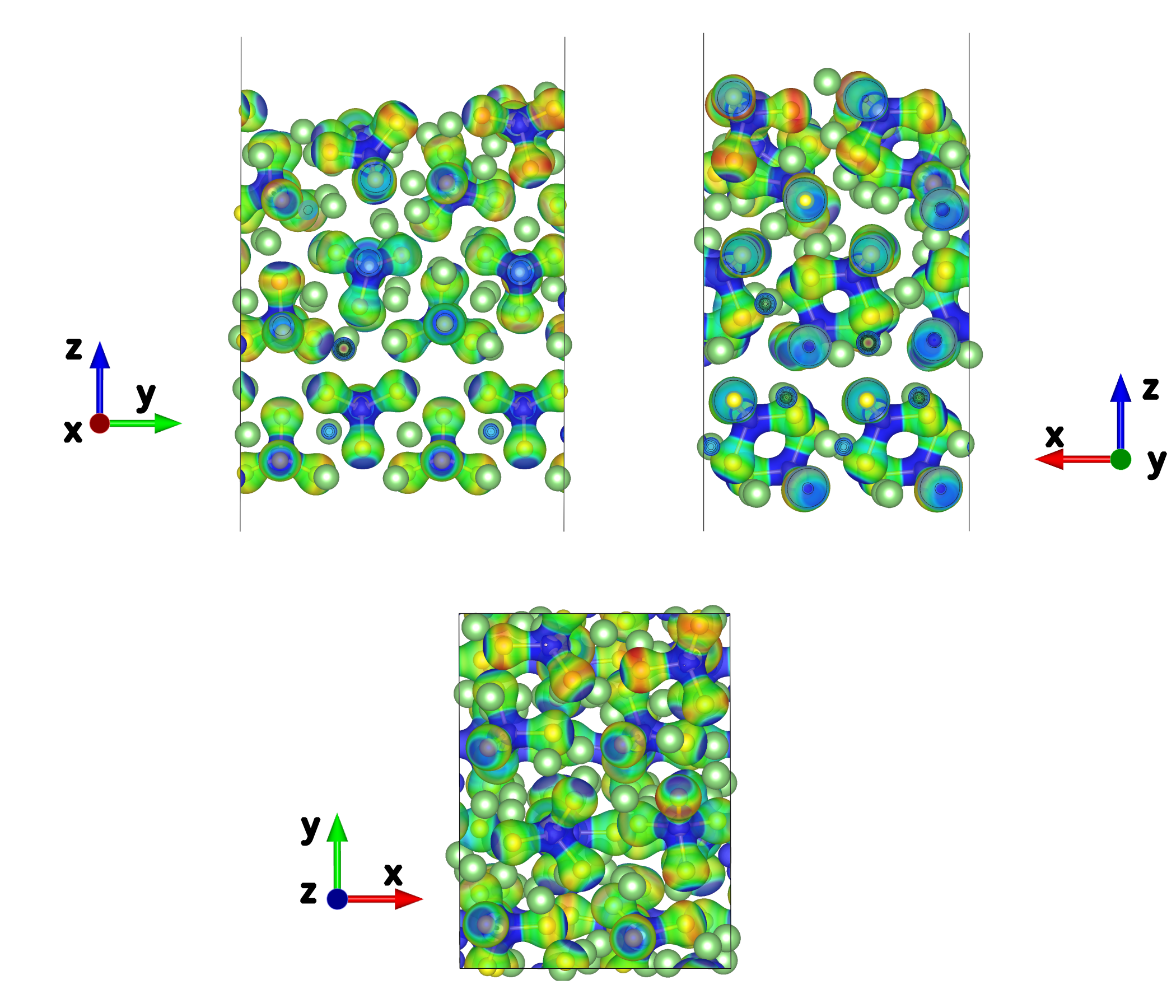} \\
  \caption{Electrostatic potential of the MD snapshot after 50~$ps$.}
  \label{fig:SImd500}
\end{figure*}

\begin{figure*}[ht]
  \centering
    \includegraphics[width=0.7\linewidth]{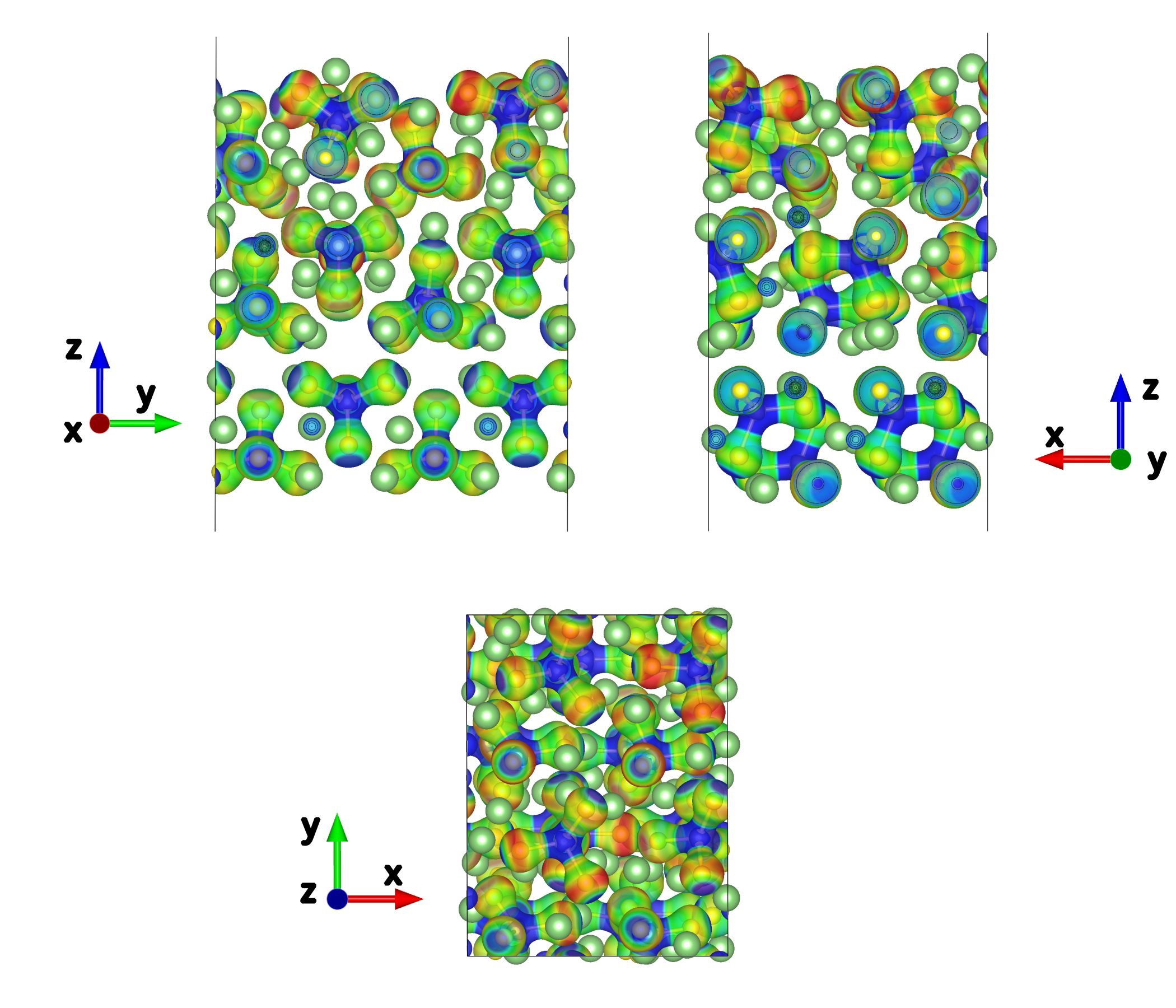} \\
  \caption{Electrostatic potential of the MD snapshot after 1000~$ps$. }
  \label{fig:SImd10000}
\end{figure*}

\begin{figure*}[ht]
  \centering
    \includegraphics[width=0.7\linewidth]{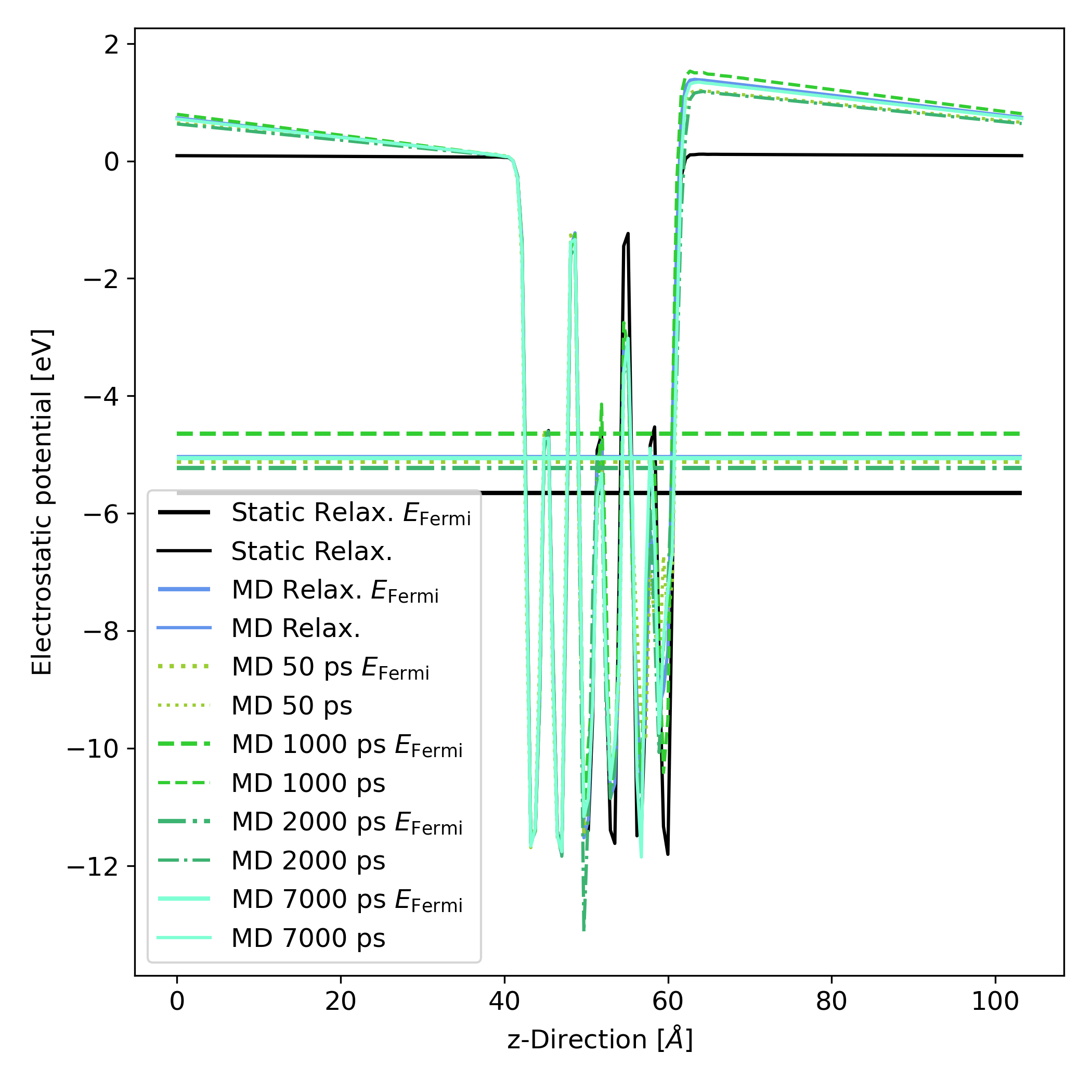} \\
  \caption{Electrostatic potential aligned to the vacuum energy of the fixed surface (at 40~\r{A}).}
  \label{fig:electrostatalign}
\end{figure*}

\begin{figure*}[ht]
  \centering
    \includegraphics{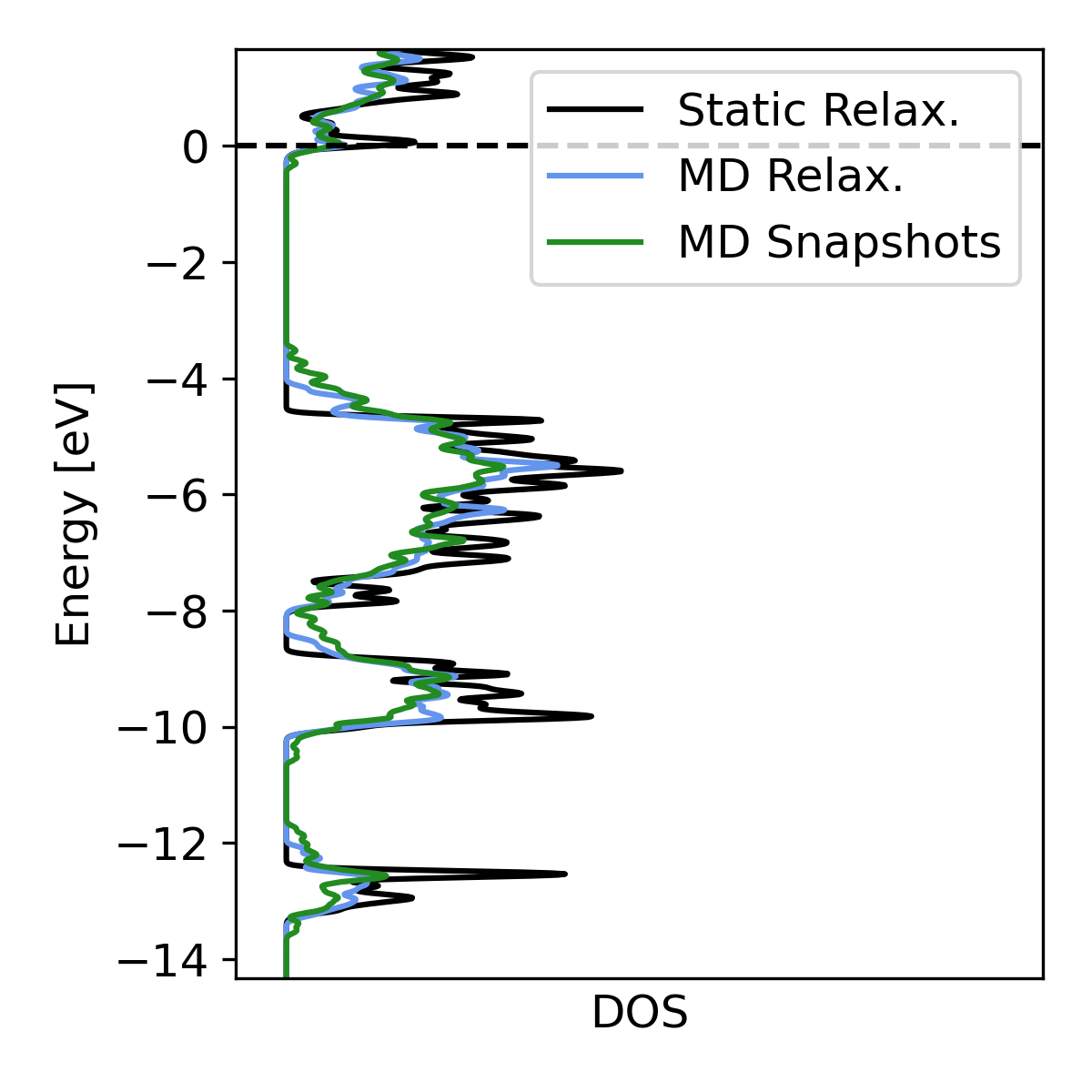} \\
  \caption{DOS according to alignment from the electrostatic potential. }
  \label{fig:dosalignment}
\end{figure*}

\end{document}